\RequirePackage{amsmath}
\documentclass[namedreferences]{solarphysics}

\usepackage{savesym}

\usepackage[hyperref,optionalrh, solaromanenum]{spr-sola-addons} 
\usepackage[dvipsnames]{xcolor}
\usepackage{graphicx}                    

\usepackage{color}                       

\usepackage{booktabs}

\usepackage{multirow}

\begin{document}

\begin{article}

\begin{opening}

\title{Solar Flare Prediction and Feature Selection using Light Gradient Boosting Machine Algorithm\\ {\it Solar Physics}}

\author[addressref={aff1,aff2},corref,email={vysakh.official1999@gmail.com}]{\inits{P.A.}\fnm{P A}~\lnm{Vysakh}\orcid{0009-0008-3146-287X}}

\author[addressref={aff3},corref]{\inits{P.}\fnm{Prateek}~\lnm{Mayank}\orcid{0000-0001-8265-6254}}

\address[id=aff1]{Anton Pannekoek Institute for Astronomy, University of Amsterdam, Netherlands}
\address[id=aff2]{Department of Physics, Indian Institute of Technology Dhanbad, India}
\address[id=aff3]{Department of Astronomy, Astrophysics and Space Engineering, Indian Institute of Technology Indore, India}

\begin{abstract}
Solar flares are among the most severe space weather phenomena, and they have the capacity to generate radiation storms and radio disruptions on Earth. The accurate prediction of solar flare events remains a significant challenge, requiring continuous monitoring and identification of specific features that can aid in forecasting this phenomenon, particularly for different classes of solar flares. In this study, we aim to forecast C and M class solar flares utilising a machine-learning algorithm, namely the Light Gradient Boosting Machine. We have utilised a dataset spanning 9 years, obtained from the Space-weather Helioseismic and Magnetic Imager Active Region Patches (SHARP), with a temporal resolution of 1 hour. A total of 37 flare features were considered in our analysis, comprising of 25 active region parameters and 12 flare history features. To address the issue of class imbalance in solar flare data, we employed the Synthetic Minority Oversampling Technique (SMOTE). We used two labeling approaches in our study: a fixed 24-hour window label and a varying window that considers the changing nature of solar activity. Then, the developed machine learning algorithm was trained and tested using forecast verification metrics, with an emphasis on evaluating the true skill statistic (TSS). Furthermore, we implemented a feature selection algorithm to determine the most significant features from the pool of 37 features that could distinguish between flaring and non-flaring active regions. We found that utilising a limited set of useful features resulted in improved prediction performance. For the 24-hour prediction window, we achieved a TSS of 0.63 (0.69) and accuracy of 0.90 (0.97) for $\geq$C ($\geq$M) class solar flares.
\end{abstract}

%

\end{opening}

%
\section{Introduction}

The study of the Sun's influence on Earth and space, commonly referred to as space weather, has become a crucial field of research worldwide. It involves investigating the impact of changes in the Sun's magnetic field on Earth's magnetic field and upper atmosphere, leading to geomagnetically induced disturbances. Solar flares, coronal mass ejections (CMEs), solar energetic particles, and solar wind stream interaction regions (SIRs) are considered to be the major space weather drivers. These space weather events can cause disturbances in various technological systems, such as electric power supply, navigation, and satellite functionality, and even pose health hazards to astronauts. Additionally, extreme space weather events may result in significant economic losses \citep{oughton_2019_a}. Due to the risks posed by such events, space weather forecasting is crucial for preparing a rational response to mitigate their impact.\\

Among the various drivers of space weather, solar flares hold a distinctive position due to their rapid travel time, their electromagnetic signal taking only about 8 minutes to reach Earth's location. Additionally, it has been observed that M-class solar flares can result in small radiation storms and brief radio interruptions, particularly in the polar regions \citep{ECHER2005855}. X-class solar flares, which are more powerful than M-class flares, can cause radiation storms with extended durations and more severe impacts. In contrast, CMEs typically take around 1-3 days, and SIRs take around 3-4 days to reach Earth. Several forecasting methods have been developed for predicting CMEs and SIRs, including probabilistic approaches (e.g., PDF \citep{bussyvirat_2014_predictions}; PROJECTZED \citep{riley_2017_forecasting}; AnEn \citep{owens_2017_probabilistic}), observation-based empirical approaches (e.g., ESWF \citep{reiss_2016_verification}; WSA \citep{arge_2000_improvement}), and magnetohydrodynamic approaches (e.g., MAS \citep{riley_2001_an}; ENLIL \citep{dodstrcil_2003_modeling}; SWMF \citep{gbortth_2011_obtaining}; EUHFORIA \citep{jenspomoell_2018_euhforia}; SWASTi \citep{mayank_2022_swastisw}). However, when it comes to predicting solar flares, probabilistic methods stand out as the only efficient option. This is due to their ability to provide results promptly, which is crucial considering the rapid travel time of solar flares.\\

Over the past decade, significant progress has been made in probabilistic methods for solar flare prediction, leading to the emergence of various techniques. \cite{huang_2013_solar, nishizuka_2017_solar} utilised the Decision Tree algorithm in their solar flare model, while \cite{li_2013_solar} employed Neural Networks and Learning Vector Quantisation models. The Support Vector Machine algorithm has been widely employed in solar flare prediction, with studies conducted by \cite{bobra_2015_solar, nishizuka_2017_solar, florios_2018_forecasting, ribeiro_2021_machine}. The other machine learning based solar flare model includes the k-NN method \citep{nishizuka_2017_solar}, Multi-Models \citep{liu_2017_shortterm}, the Random Forest \citep{liu_2017_shortterm, florios_2018_forecasting, ribeiro_2021_machine}, Multi-layer Perceptrons \citep{florios_2018_forecasting}, Long-Short Term Memory (LSTM) \citep{liu_2019_predicting, wang_2020_predicting, jiao_2020_solar, chen_2019_identifying}. Additionally, Deep Learning Neural Networks have been used for solar flare prediction \citep{nishizuka_2018_deep}. Despite substantial advancements in diverse prediction models for solar flares, a definitive consensus regarding the optimal performing model is currently lacking.\\

Choosing reliable features for solar flare prediction is a crucial step in achieving accurate results. Many prediction models rely on photospheric magnetic field data to parameterise active regions (ARs) and describe them using a few key parameters, with the goal of establishing relationships between the behaviour of the photospheric magnetic field and solar activity. However, there is significant variability in the AR parameters considered in these models. Some models focus on characterising the magnetic field topology of ARs \citep[e.g.,][]{schrijver_2007_a}, while others measure the integrated Lorentz force exerted by an AR \citep[e.g.,][]{fisher_2011_global}, or employ parameterisations for energy, helicity, currents, and shear angles \citep[e.g.,][]{moore_2012_the, labonte_2007_survey, leka_2003_photospheric}.\\

In this study, we have developed a solar flare prediction model utilising the Light Gradient Boosting Machine \citep[LightGBM;][]{ke_2017_lightgbm} algorithm, and have conducted a comprehensive comparison of various active region (AR) features and flaring history parameters to determine their effectiveness in achieving accurate forecasts. LightGBM, a classification method, has received limited attention in the context of solar flare prediction, with only a few studies, including \cite{ribeiro_2021_machine}, investigating its performance in comparison to Support Vector Machine (SVM) and Random Forest (RF) algorithms.\\

The paper is broadly divided into four sections. Section \ref{sec:DCP} explains the data used and how it was processed for the machine learning algorithm to take it as input. Section \ref{sec:MLC} explains in detail the machine learning model used and the improvements, as well as the metrics used to study the performance of the model. Section \ref{sec:RnD} discusses the results obtained and explains the feature selection algorithm used to identify the best set of features. Additionally, a comprehensive comparison with similar models is also presented. Finally, Section \ref{sec:SnC} concludes the paper by discussing both the improvements and the limitations of the methods used.

\section{Data Collection and Preparation }\label{sec:DCP}
\subsection{SHARP HMI active region parameters}
For our model, the dataset we adopted was the  Space-weather HMI Active Region Patches \citep[SHARP,][]{Bobra_2014}) data provided by SDO HMI (Solar Dynamics Observatory: Helioseismic and Magnetic Imager). In 2014, the SHARP data series was released, which includes maps in patches that cover automatically tracked magnetic concentrations throughout their lifetime \citep{Bobra_2014}. These patches identify the active regions (ARs) automatically and continuously calculate various summary parameters of ARs at a 12-minute cadence. We collected a total of 25 AR summary parameters provided by the SHARP module. The selection of the 25 parameters was based on the previous work done in \cite{Bobra_2014}. The list of initially considered parameters characterises several features of solar active regions that have been linked to increased flare activity. These parameters include different types of indices, such as the total magnetic flux, the spatial field gradients, the vertical current density characteristics, current helicity, and a proxy for integrated free magnetic energy \citep{Bobra_2014}. Table \ref{tbl:TableSHARP} presents the initial list of parameters along with their corresponding equations and a brief description. These parameters are consistent with those used in \cite{Bobra_2014}. We used SunPy python module \citep{Community_2015} to collect SHARP data using JSOC client. The data products used were \emph{hmi.sharp.720s} and \emph{cgem.lorentz}. The subsequent data series, as described in \cite{sun2019cgem}, estimates the Lorentz force in ARs based on vector magnetogram patches from the HMI. Both the SHARP and HMI data series were collected at a cadence of 1 hour.

\subsection{GOES Flaring data}
To define the training samples for the machine learning model, we searched for flares that occurred between January 2012 and December 2020. This time range extends the whole range in which the SHARP HMI data is available. We used SunPy module to collect flaring events from the Geostationary Operational Environmental Satellite (GOES) Xray Flux Catalog maintained by the National Centers for Environmental Information. In order to create the final dataset, each active region was uniquely identified by its NOAA AR number. This identifier facilitated the cross-matching of GOES data with HMI SHARP data. Specifically, for every event listed in the GOES catalogue, all observations corresponding to the same NOAA AR number were extracted from the HMI SHARP dataset. From this subset of observations, the one closest in time to the occurrence of the solar flare was designated as a positive event. This process was repeated for each flare within the GOES catalogue, thereby culminating in the formation of the comprehensive final dataset. It is important to acknowledge that a small subset of events documented in the GOES catalogue was not present in the HMI SHARP catalog. Consequently, these observations were omitted during the assembly of the final dataset.  After filtering out the samples without any observations in the HMI SHARP dataset, we ended up with 1846 B-Class flares, 3246 C-Class flares, 318 M-Class flares, and 23 X-Class flares. It should be noted that we consider all the flaring events irrespective of the AR it originates from, i.e. if an active region flares multiple times during its lifetime, we count each of them as separate events.

\subsection{Flaring history parameters}
Active regions being areas of rich magnetic activity on the surface on the Sun, it is a good assumption that an active region with a history of flare occurrences could hold a high probability for more flares happening. 
\begin{figure}[hbt!]
    \centering
    \includegraphics[width=\textwidth]{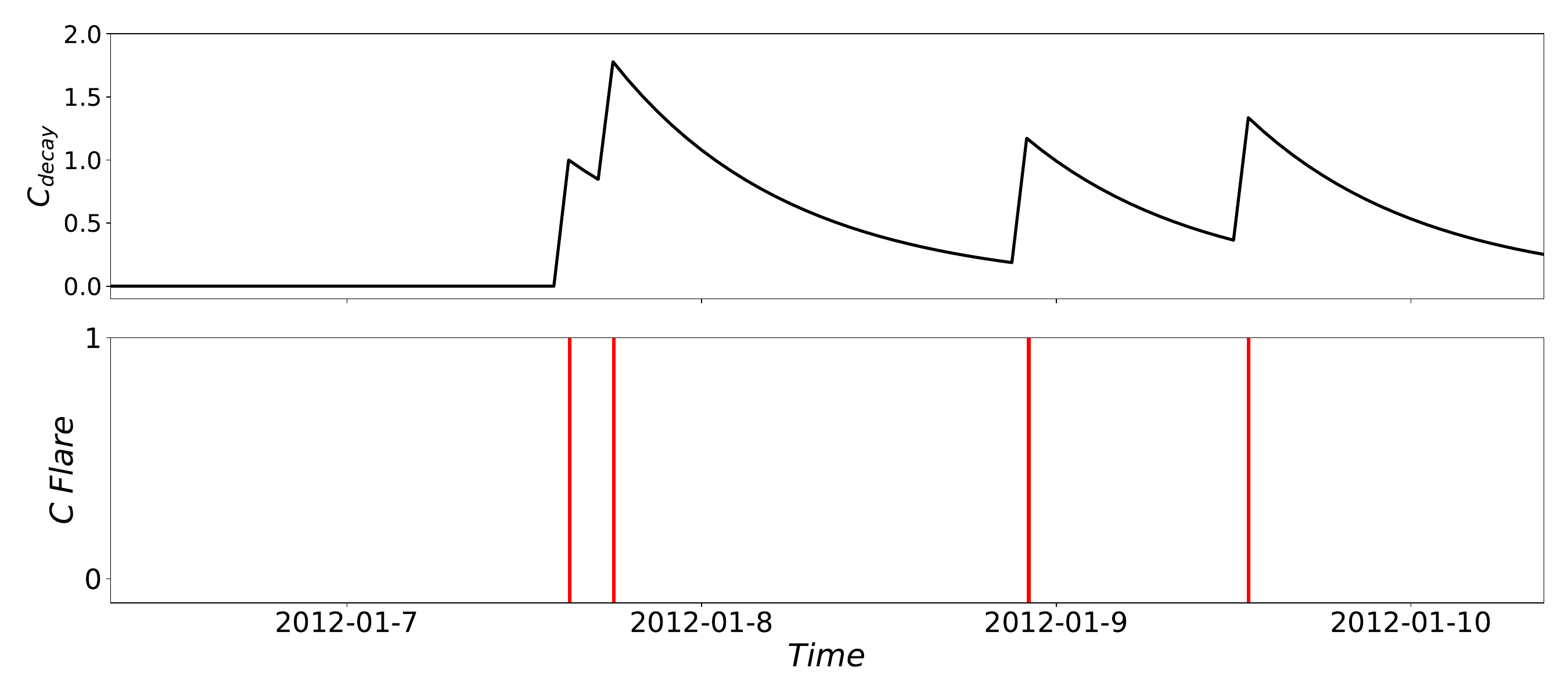}
    \caption{The plot of C$_{decay}$ as a function of time. Comparing the top and bottom panels, we can see that when a flaring event occurs, the value increases and falls exponentially until the active region produces another flare. This plot corresponds to the NOAA AR 11393.}
    \label{fig:Cdec}
\end{figure} 
Thus, through understanding the past flaring activity of an active region, we should be able to increase the ability to forecast flares \citep{Falconer_2012}. Based on the collected flaring data, we made a framework of 12 flaring history parameters \citep{liu_2019_predicting} consisting of flare decay values and previous flare occurrences. These parameters will be able to encompass the history of each active region in terms of various levels of flaring activity, thus enabling us to study the contribution of history parameters in detail. Using the equations given in \cite{Jonas_2018}, time decay values were calculated for each individual active region. For a data sample in an AR, the decay values w.r.t C-class, M-class, and X-class are given by:

\begin{equation}
    \centering
    \mathrm{C_{decay}(t)}=\sum_{f_i\in F_C}exp\Big(-\frac{t-t(f_i)}{\tau}\Big)   \quad,
\end{equation}
\begin{equation}
    \centering
    \mathrm{M_{decay}(t)}=\sum_{f_i\in F_M}exp\Big(-\frac{t-t(f_i)}{\tau}\Big)   \quad,
\end{equation}
\begin{equation}
    \centering
    \mathrm{X_{decay}(t)}=\sum_{f_i\in F_X}exp\Big(-\frac{t-t(f_i)}{\tau}\Big)   \quad,
\end{equation}
here $F=F_C \cup F_M \cup F_X$,where $F_k$ corresponds to the set of $k$ class flares, and $t(f_i)$ corresponds to the time of occurrence of the flare $f_i$. The value of $\tau$ is fixed to be 12 as proposed by \cite{Jonas_2018}. The nature of this decay function is depicted in Figure \ref{fig:Cdec} with $C_{decay}$ values of an active region. Apart from the time decay values w.r.t flare classes, we also calculate the energy decay values of an AR considering all flares, regardless of the type of flare that occurred before the sample time t. Here $E(f_i)$ corresponds to the magnitude of the flare $f_i$.
\begin{equation}
    \centering
    \mathrm{E_{dec}(x_t)}=\sum_{f_i\in F}E(f_i) exp\Big(-\frac{t-t(f_i)}{\tau}\Big)   \quad,
\end{equation}
\begin{equation}
    \centering
    \mathrm{logE_{dec}(x_t)}=\sum_{f_i\in F}\log(E(f_i)) exp\Big(-\frac{t-t(f_i)}{\tau}\Big)   \quad.
\end{equation}

The remaining seven parameters include flare history features of a data sample as described in \cite{nishizuka_2017_solar}. These features include Chist (Mhist, Xhist) storing the total number of C-Class (M-Class, X-Class) flares in the AR before the observation time, Chist1d (Mhist1d, Xhist1d) storing the flaring activity of each class during the 24 hours prior observation time, and finally Xmax1d storing maximum flare intensity 24 hours prior to the observation. Table \ref{tbl:flarehistParams} summarises all the flaring history parameters adopted in this model. As of now, we have 25 active region parameters and 12 flaring history parameters making a total of 37 features. 

\begin{figure}[hbt!]
    \centering
    \includegraphics[width=120mm]{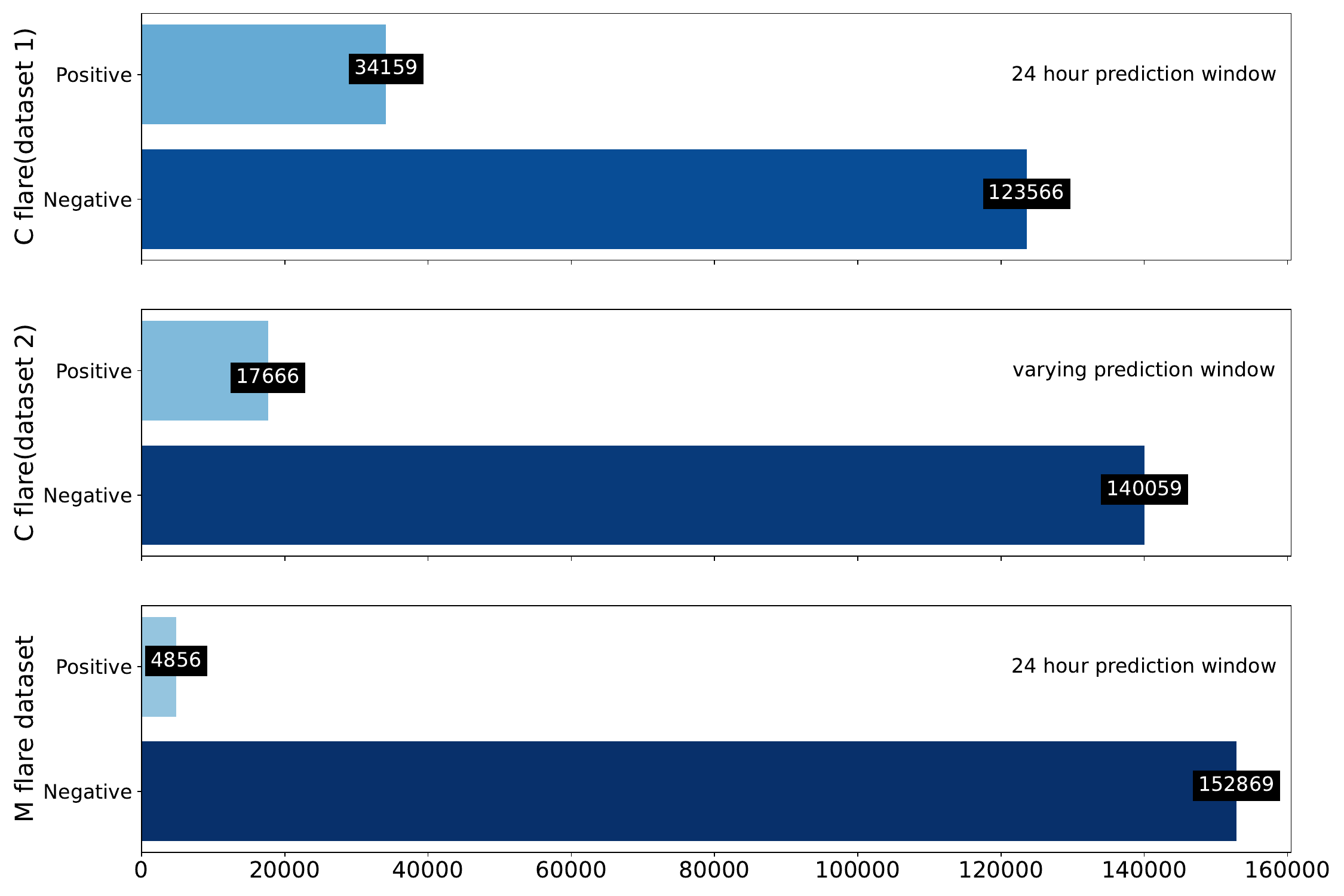}
    \caption{Positive and negative classes in 3 different datasets ($\geq$C-Class with 24 hour prediction window, $\geq$C-Class with varying prediction window, and $\geq$M-Class with 24 hour prediction window).}
    \label{fig:imbalance}
\end{figure}
\begin{table}[hbt!]

    \centering 
    
    \makebox[\textwidth]{%
    \caption{SHARP active region parameters formula and description}
    \label{tbl:TableSHARP}
    
    \begin{tabular}[width=\textwidth]{l l r} 
        \hline\hline 
        Keyword & Formula & Description\\ [0.5ex] 
        \hline 
        TOTUSJH & $H_{c_{total}}\propto \sum |B_z.J_z|$ & Total unsigned current helicity \\
        TOTUSJZ & $J_{z_{total}}= \sum |J_z|dA$ & Total unsigned vertical current \\
        USFLUX & $\Phi=\sum |B_z|dA$  & Total unsigned flux \\
        AREA\_ACR & Area = $\sum $Pixels & Area of strong field pixels in the active region \\
        SAVNCPP & $J_{z_{sum}}\propto \left| \sum^{B^+_z}J_zdA \right |+\left|\sum^{B^-_z}J_zdA \right|$ & Sum of the absolute values of the net current per polarity \\
        TOTPOT & $\rho_{tot}\propto \sum \left( \vec{B}^{Obs}-\vec{B}^{Pot} \right)^2 dA$  & Total photospheric magnetic free energy density \\
        R\_VALUE & $\Phi =\sum |B_{LoS}|$ dA within R mask  & Sum of flux near polarity inversion line \\
        TOTUBSQ & $F\propto \sum B^2$ & Total magnitude of Lorentz force \\
        ABSNJZH & $H_{c_{abs}}\propto |\sum B_z.J_z|$ & Absolute value of the net current helicity \\
        MEANPOT & $\overline{\rho}\propto \frac{1}{N}\sum \left( \vec{B}^{Obs}-\vec{B}^{Pot}\right)^2$ & Mean photospheric magnetic free energy \\
        MEANSHR & $\overline{\Gamma} = \frac{1}{N}\sum arccos\left(\frac{\vec{B}^{Obs}.\vec{B}^{Pot}}{|B^{Obs}||B^{Pot}|}\right) $ & Mean shear angle \\
        SHRGT45 & Area with Shear $>45^{\circ}$ / Total Area & Area fraction with a shear angle higher than $45^{\circ}$ \\
        TOTFZ & $F_z\propto \sum \left( B^2_x+B^2_y-B^2_z\right)dA$ & Sum of z-component of Lorentz force \\
        MEANGAM & $\overline{\gamma} = \frac{1}{N}\sum arctan\left(\frac{B_h}{B_z}\right)$ & Mean angle of field from radial \\
        TOTFY & $F_y\propto \sum B_yB_zdA$ & Sum of y-component of Lorentz force \\
        MEANGBT & $\overline{|\nabla B_{tot}|}=\frac{1}{N}\sum \sqrt{\left(\frac{\partial B}{\partial x}\right)^2+\left(\frac{\partial B}{\partial y}\right)^2}$ & Mean gradient of total field \\
        MEANGBH & $\overline{|\nabla B_h|}=\frac{1}{N}\sum \sqrt{\left(\frac{\partial B_h}{\partial x}\right)^2+\left(\frac{\partial B_h}{\partial y}\right)^2}$ & Mean gradient of horizontal field \\
        MEANJZD & $\overline{J_z}\propto \frac{1}{N}]\sum \left( \frac{\partial B_y}{\partial x}-\frac{\partial B_x}{\partial y}\right)$ & Mean vertical current density \\
        MEANGBZ & $\overline{|\nabla B_z|}=\frac{1}{N}\sum \sqrt{\left(\frac{\partial B_z}{\partial x}\right)^2+\left(\frac{\partial B_z}{\partial y}\right)^2}$ & Mean gradient of vertical field \\
        EPSY & $\partial F_y\propto \frac{-\sum B_yB_z}{\sum B^2}$ & Sum of y-component of normalised Lorentz force \\ 
        EPSX & $\partial F_x\propto \frac{\sum B_xB_z}{\sum B^2}$ & Sum of x-component of normalised Lorentz force \\
        TOTFX & $F_x\propto -\sum B_xB_zdA$ & Sum of x-component of Lorentz force \\
        EPSZ & $\partial F_z\propto \frac{\sum B_x^2+B_y^2-B_z^2}{\sum B^2}$ & Sum of z-component of normalised Lorentz force \\       
        MEANALP & $\alpha_{total}\propto \frac{\sum J_z . B_z}{\sum B^2_z}$ & Mean characteristic twist parameter, $\alpha$ \\
        MEANJZH & $\overline{H_c}\propto \frac{1}{N}\sum B_z . J_z$ & Mean current helicity \\
        \hline 
    \end{tabular}
    }
    
\end{table}

\begin{table}[hbt!]
    \centering
   
    \makebox[\textwidth]{%
    
    \begin{tabular}[b]{l l r}
        \hline\hline 
        Keyword & Formula & Description\\ [0.5ex] 
        \midrule
        Cdecay & Cdecay($x_t$)= $\sum_{f_i\in F_C}e^{-\frac{t-t(f_i)}{\tau}}$ & Time decay value based on the \\ & & past C-class flares \\
        Mdecay & Mdecay($x_t$)= $\sum_{f_i\in F_M}e^{-\frac{t-t(f_i)}{\tau}}$ & Time decay value based on the \\ & & past M-class flares \\
        Xdecay & Xdecay($x_t$)= $\sum_{f_i\in F_X}e^{-\frac{t-t(f_i)}{\tau}}$ & Time decay value based on the \\ & & past X-class flares \\
        Edecay & Edecay($x_t$)= $\sum_{f_i\in F}E(f_i) e^{-\frac{t-t(f_i)}{\tau}}$ & Time decay value determined using magnitudes \\ & & of all past flares \\
        logEdec & logEdec($x_t$)= $\sum_{f_i\in F}\log(E(f_i)) e^{-\frac{t-t(f_i)}{\tau}}$ & Time decay value determined using log-magnitudes \\ & &of all past flares \\
        Chist & - & Total number of C-class flares ever recorded in an AR \\
        Mhist & - & Total number of M-class flares ever recorded in an AR \\
        Xhist & - & Total number of X-class flares ever recorded in an AR \\
        Chist1d & - & C-class flare activity in an AR over 24 hours \\
        Mhist1d & - & M-class flare activity in an AR over 24 hours \\
        Xhist1d & - & X-class flare activity in an AR over 24 hours \\
        Xmax1d & - & Maximum X-ray intensity 24 hours before \\
         \bottomrule 
    \end{tabular}
    }
    \caption{Flaring history parameters formulae and description}
    \label{tbl:flarehistParams}
\end{table}

\subsection{Class imbalance problem}
To obtain valuable insights into the robust predictive performance of the data, it is important to ensure the integrity of the data \citep{ahmadzadeh2019challenges}. As we are taking the approach of point-in-time prediction, according to the study by \cite{ahmadzadeh2019challenges}, one of the major factors that affect the performance of the model is the extensive class imbalance between positive and negative samples.\\

Figure \ref{fig:imbalance} shows the positive and negative samples in each dataset prepared. The figure clearly shows that the number of negatively sampled events far outnumber the positively sampled events. This imbalance becomes much more significant at datasets considering higher flare magnitudes. If this imbalance is not properly dealt with, the results and predictions can be misleading as the data is biased towards non-flaring events. There are various methods to deal with imbalanced problems. In this work, we use an oversampling technique, namely,  Synthetic Minority Over-sampling Technique (SMOTE). Normal oversampling techniques randomly duplicate examples in minority class without adding any new information to the set. SMOTE, as described in \cite{Chawla_2002}, works by selecting examples that are close in feature space and interpolating them to find new examples.\\

During the course of time, there were several modifications to the basic SMOTE algorithm changing the approach on how each selects samples from the minority class. A few of the most commonly used are Borderline SMOTE \citep{10.1007/11538059_91}, Borderline
SMOTE SVM \citep{nguyen2011borderline} and Adaptive Synthetic Sampling \citep{4633969}. Trial runs of the whole flare prediction algorithm with these three variations were conducted separately, in which the Borderline SMOTE algorithm performed the best. Thus we have used the Borderline SMOTE method along with random undersampling \citep{Chawla_2002} to create the final required dataset. This will bring the ratio of flaring events to non-flaring events to a value around 0.6. This value was chosen by running a grid search on values from 0.1 to 0.9 and optimising the performance of the classifier.

\subsection{Standardisation}
As the features are of very different scales, the model could give higher preference to those features with higher numerical values. To prevent this, we standardise the values of each feature, bringing them to a comparable scale.

The following equation is used to calculate the standardised values for each feature. For $n^{th}$ feature of the $m^{th}$ data sample, the standardised value is given by:
\begin{equation}
    \centering
    s^m_n = \frac{v^m_n-\mu_n}{\sigma_n}   \quad,
\end{equation}
where $v^m_n$ is the actual value of the $n^{th}$ feature, and $\mu_n$ and $\sigma_n$ are the mean and standard deviation of the $n^{th}$ feature respectively.

\section{Machine Learning Classifier}\label{sec:MLC}
\subsection{Labeling Algorithm}\label{sec:MLC1}
\begin{figure}[hbt!]
    \centering
    \includegraphics[width=\textwidth]{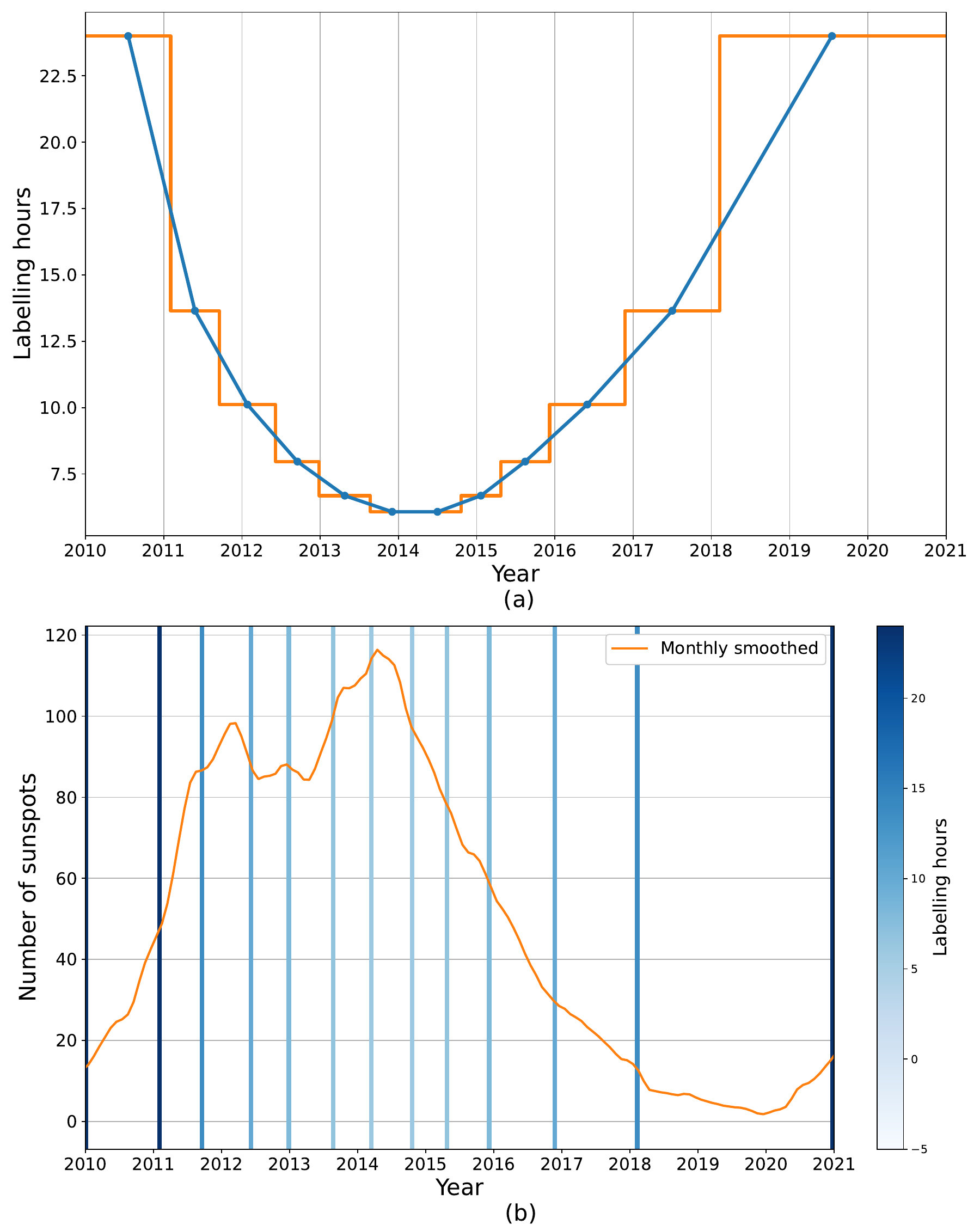}
    \caption{(a) Varying label prediction window (b) Divisions made throughout the solar cycle with color map displaying the prediction window.}
    \label{fig:labeling_int}
\end{figure}

As our prediction model is a binary classifier, we define the flaring and non-flaring events as binary flags. First, we used the operational form of associations between active regions and solar flares to label the flares. While using the operational form, if an active region produces any flares within 24 hours after the sample time, it is classified as a positive class or flaring event, and if it does not flare within the given time, it is classified as a negative class or the non-flaring event. Secondly, we designed a new labelling algorithm taking into account the varying nature of solar activity throughout the cycle. The new algorithm divides the solar cycle into 12 parts, keeping flare occurrences equal in each and allotting each division a prediction window which varies logarithmically from 6 hours to 24 hours in the active and quiet periods of the solar cycle. Figure \ref{fig:labeling_int}(a) and Figure \ref{fig:labeling_int}(b) show how the prediction window varies and how it is allotted throughout the solar cycle.\\

For our model, we will be  considering C-Class, M-Class, and X-Class flares as positive events and the rest as negative. Furthermore, in terms of flare magnitude, we prepared two datasets such that the first one considers any flare with a magnitude greater than or equal to a C-Class flare to be positive and the second one such that any flare with a magnitude greater than or equal to an M-Class flare is considered positive. Due to high-class imbalance in the $\geq$M-Class dataset, only the operational form of labeling is adopted for the same. Thus we have three datasets: $\geq$C-Class dataset (24 hour prediction window), $\geq$C-Class dataset (varying prediction window) and $\geq$M-Class dataset (24 hour prediction window).

\subsection{Light Gradient Boosting Machine Classifier} \label{subsec:lightgbm}
We use a LightGBM model as our classifier. LightGBM \citep{ke2017lightgbm} is a highly efficient Gradient boosting decision tree (GBDT) algorithm developed by Microsoft Corporation in 2016. Compared to other GBDT algorithms like XGboost, LightGBM uses an optimised histogram-building method by down-sampling data and features using GOSS (Gradient Based One Side Sampling) and EFB (Exclusive Feature Bundling).\\

\begin{figure}[hbt!]
    \centering
    \includegraphics[width=0.9\textwidth]{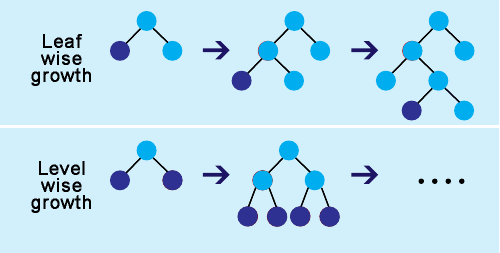}
    \caption{Leaf-wise and level-wise growths.}
    \label{fig:leaflevel}
\end{figure}
In contrast to its gradient-boosting counterparts, LightGBM's histogram building approach results in significant improvements in efficiency and forecasting ability. The algorithm discovers complex data patterns by concentrating on expanding leaf nodes that have the greatest impact on loss reduction, which is a significant advantage for our solar flare prediction task using HMI SHARP data. This choice of classifier perfectly complements the complex nature of solar flare predictors and triggers in our dataset. LightGBM is equipped to understand intricate relationships and distinctions thanks to the leaf-wise growth technique, which effectively decodes complex solar phenomena. Hyperparameters are used to further fine-tune the algorithm's performance, achieving a balance between computational needs and predictive power across various dataset sizes. Nevertheless, it is important to recognise the hindrances of the traditional level-wise growth approach from which LightGBM deviates. The primary distinction between leaf-wise and level-wise growth lies in how decision trees are constructed. In the leaf-wise approach, individual leaves at the end of the tree are allowed to grow independently, as opposed to the level-wise method, where growth occurs level by level sequentially. This difference has been visualised in Figure \ref{fig:leaflevel}. Even though level-wise growth ensures a balanced tree structure and reduces overfitting, it might miss complex patterns in our data on solar activity. With its deeper trees and greater ability to capture complex interactions, LightGBM's adoption of the leaf-wise growth strategy presents a compelling alternative tailored to the intricacies of solar flare prediction. The appeal of LightGBM goes beyond its algorithmic innovation and includes its resource-conserving architecture. The algorithm uses gradient-based optimisation techniques like GOSS and EFB to speed up training and feature extraction, which is a significant benefit when dealing with a large dataset such as in the case of solar flare data.\\

The utilisation of LightGBM introduces a comprehensive array of hyperparameters, facilitating a nuanced configuration of the model. Within our study, the hyperparameters subject to tuning encompass the bagging fraction, learning rate, number of leaves, L1 regularisation, and L2 regularisation (\citealp{ke2017lightgbm}). The bagging fraction assumes a pivotal role by governing the proportion of training data used in each iterative step, thereby contributing to the model's generalisation capability. Notably, the refinement of hyperparameters such as the number of leaves,  and the imposition of L1 and L2 regularisation mechanisms collectively contribute to the model's regularisation process. The optimisation of these parameters is undertaken through an exhaustive grid search method aimed at eliciting the optimal configuration that maximises model performance, gauged through the TSS metric. The ranges considered for each parameter are bagging fraction $\epsilon\; (0,1)$, learning rate $\epsilon\; (0.0001,0.1)$, number of leaves $\epsilon \;(2^4,2^8)$, L1 $\epsilon\; (0,1)$ and L2 $\epsilon\; (0,1)$. Each parameter is sampled from a uniform prior in the given range. Even though this is not the best method in terms of efficiency, we chose this approach as it had a straightforward implementation and we were able to complete it within a reasonable time frame.  \\

One of the parameters used in training the model is the loss function. We have used the focal loss as the chosen loss function for our model. As per \cite{Lin_2017_ICCV}, the focal loss is a better choice for imbalanced problems than metrics like cross-entropy, as the former assigns more weight on easily misclassified labels and down-weight easily classified labels. For a binary classification problem, if $y\in\{\pm1\}$ is the ground truth class and $p\in[0,1]$ is the model's estimated probability for class $y=1$, We could write the $\alpha$-balanced variant of focal loss to be,
\begin{equation}
    \centering
    \mathrm{FL}(p_t)=-\alpha(1-p_t)^{\gamma}\log(p_t)   \quad,
\end{equation}
\begin{equation}
    \centering
    \mathrm{where,}\; p_t = \left\{
        \begin{array}{ll}
            p & \mathrm{if}\;y=1 \\
            1-p & \mathrm{otherwise}   \quad,
        \end{array}
    \right. 
\end{equation}
where $t$ indicates the true class. This loss function is minimised over iterations to train the classification model. Optimising focusing parameter $\gamma$ and weighing parameter $\alpha$ lets the model perform better by assigning more weight to easily misclassified parameters. We set early stopping rounds parameter to be 20, to let the model stop training after a certain score in the validation set has been reached so as to avoid over-fitting. 
\subsection{Performance metrics}
\begin{figure}[hbt!]
    \centering
    \includegraphics[width=0.6\textwidth]{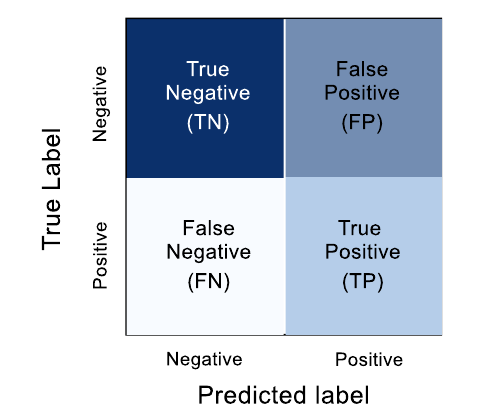}
    \caption{Confusion matrix.}
    \label{fig:confusion}
\end{figure}

Being an imbalanced binary classification problem, most common metrics like accuracy would not act as a functional measure of the performance (\cite{5128907}). This is because the high accuracy could easily be achieved by only predicting the majority class, i.e. non-flaring class. 
A confusion matrix, as shown in Figure \ref{fig:confusion}, can be used to represent the results of a binary classifier. The classifier's accurate predictions of flaring ARs are referred to as True Positives (TP), while incorrect predictions of flaring ARs are referred to as False Negatives (FN). Similarly, True negatives (TN) refer to accurate predictions of non-flaring ARs, and False positives (FP) refer to incorrect predictions of non-flaring ARs. \\

To statistically analyse the model results and compare with the observed data, we chose five metrices which are commonly used in the literature. Their expressions are following:

\begin{gather}
        \mathrm{Precision = \frac{TP}{TP+FP}}\label{eq:precision} \quad,\\ 
        \mathrm{Recall = \frac{TP}{TP+FN}}\label{eq:recall} \quad,\\
        \mathrm{HSS = \frac{2(TP \times TN - FP \times FN) }{(TP+FN)(FN+TN)+(TP+FP)(FP+TN)}}\label{eq:HSS} \quad,\\
        \mathrm{TSS = \frac{TP}{TP+FN}-\frac{FP}{TN+FP}}\label{eq:TSS} \quad,\\
        \mathrm{BACC = \frac{1}{2} \left( \frac{TP}{TP+FN} + \frac{TN}{TN+FP}\right)}\label{eq:BACC}  \quad, 
\end{gather}

\noindent where, precision summarises the number of predicted positive classes that belong to the positive class, while recall summarises how well the positive class was predicted. HSS is Heidke Skill Score and measures the fractional improvement of the forecast over the standard forecast. TSS is True Statistical Score which is defined as the difference between the recall and false alarm rate. Even though HSS is useful, it changes despite the prediction success being held constant \citep{Bloomfield_2012}, so it is suggested to use True Skill Statistic (TSS) as the measure of performance. Additionally, TSS is unbiased to class-imbalance ratio, thus making it the most useful measure for flare forecasting methods. We will also calculate the Balanced Accuracy (BACC) as it is a good measure for studying an imbalanced classification \citep{5128907}. Larger the TSS, HSS, or BACC, the better the performance of the model. Even though we will calculate all the mentioned metrics, we will focus  more on increasing the True Skill Statistic as it provides a better measure in flare forecasting.

\section{Results and Discussion}\label{sec:RnD}

\subsection{Model Evaluation}
To find the best hyperparameters for our model, we perform a grid search in the parameter space to maximise the TSS score. We run the grid search for bagging fraction, learning rate, number of leaves, L1 regularisation, and L2 regularisation. Through trial runs, the best-boosting method for our dataset was found to be the Gradient Boosting Decision Tree (GBDT). Furthermore, we determined the optimal hyperparameters for our model are as follows: a learning rate of 0.01, a bagging fraction of 0.95, 146 leaves in the decision trees, L1 regularisation set at 0.74, and L2 regularisation set at 0.23.\\

As mentioned in Section \ref{sec:MLC1}, we prepared three datasets ($\geq$C-Class (24-hour prediction window), $\geq$C-Class (varying prediction window), $\geq$M-Class (24-hour prediction window)) to train our model. Each dataset was further divided into training, validation, and test sets. To ensure the unbiased nature of performance metrics, we need to make sure that the model performance is tested on a new dataset that it has not seen before. To ensure this, we divided the datasets such that all the samples collected from January 2012 to  December 2014 were used to train the model, January 2015 to December 2016 were used as validation set to tune the model, and finally, samples from January 2017 to December 2020 were used to test the data. This division ensures the former statement and describes the real-world scenario for predicting solar flares. \\ 

\begin{figure}[hbt!]
	\centering
	\centering
	\includegraphics[width=\textwidth]{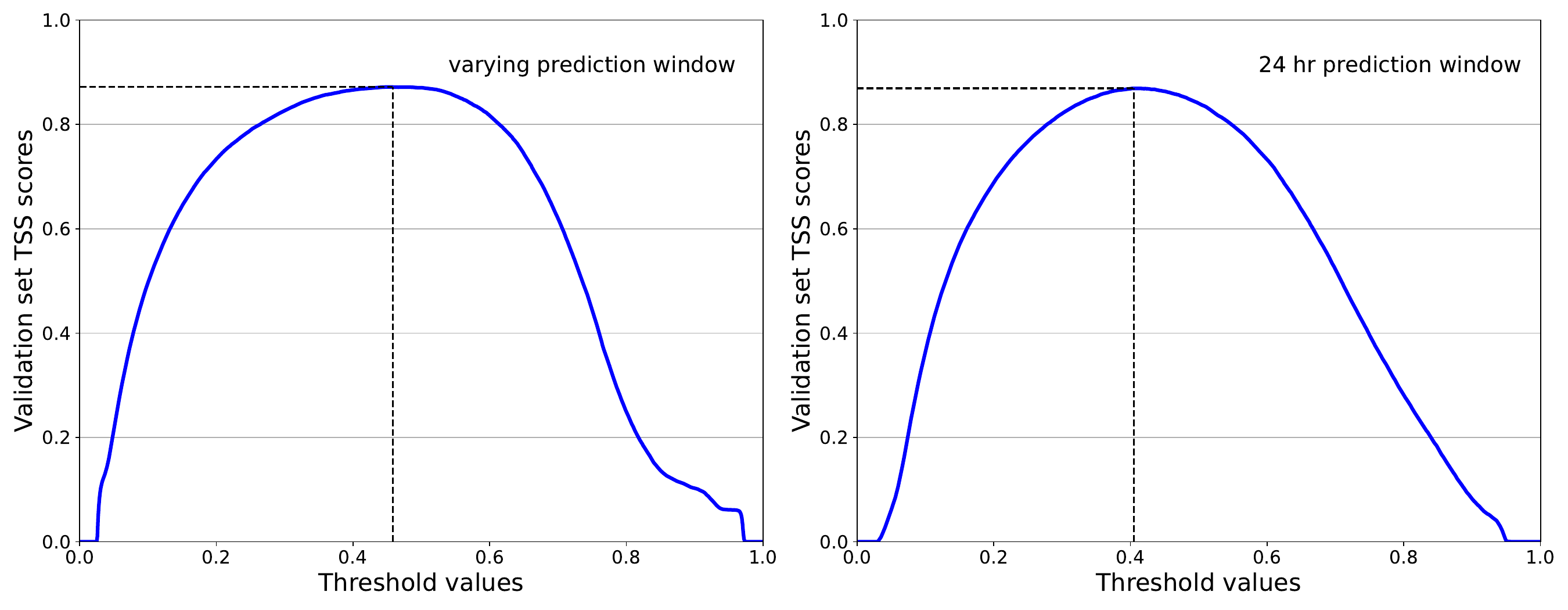}
	\caption{TSS score vs. threshold plot for a moving threshold for predicting C-class flares with varying prediction window and 24-hour prediction window.}
	\label{fig:CThreshold}
\end{figure}

\begin{figure}[hbt!]
	\centering
	\centering
	\includegraphics[width=0.6\textwidth]{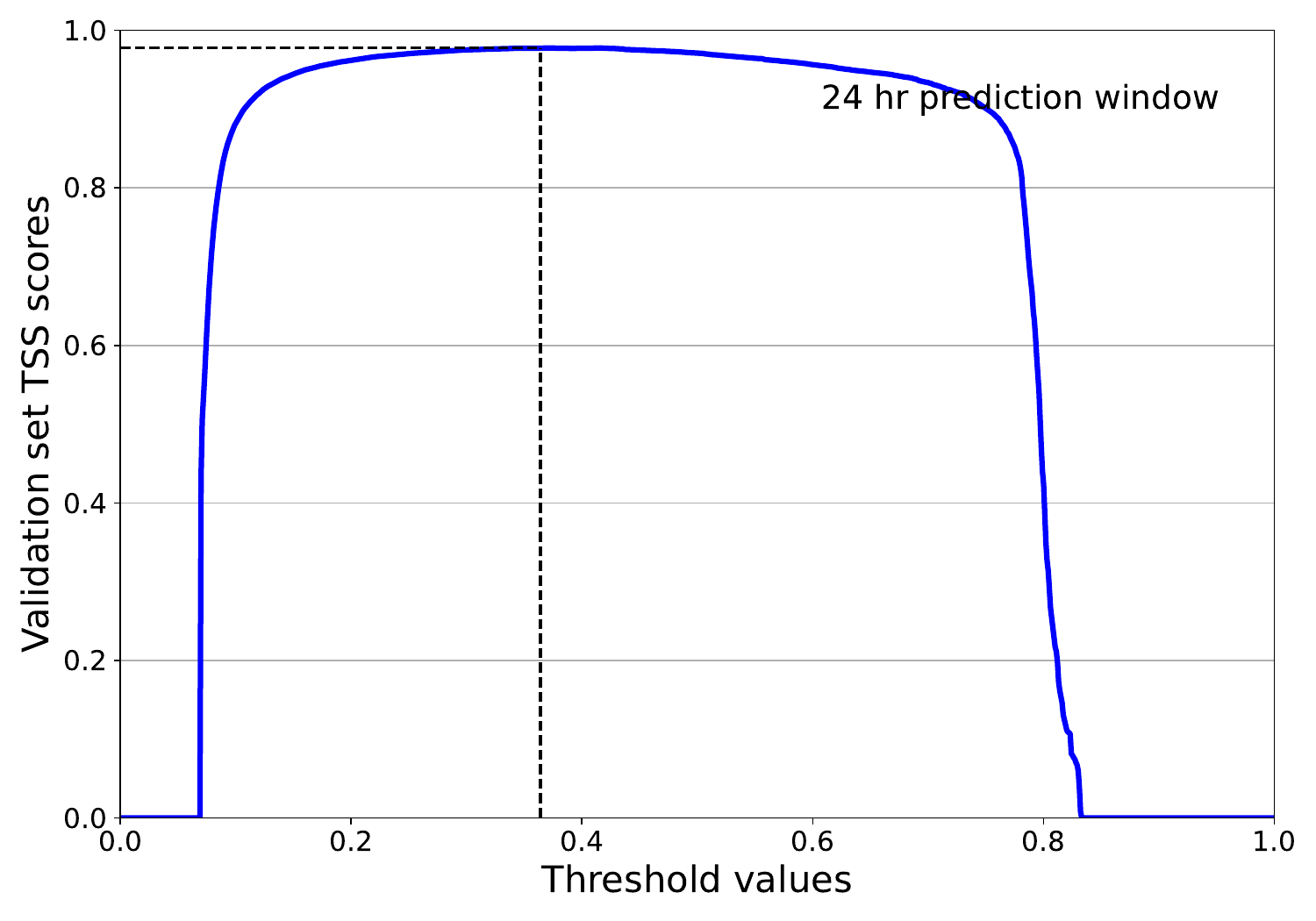}
	\caption{TSS score vs. threshold plot for moving threshold for predicting M-class flares with 24-hour prediction window.}
	\label{fig:MThreshold}
\end{figure}

Once the training, validation, and testing datasets are prepared, the LightGBM classifier is trained with the training dataset. Once the model is trained, a threshold moving approach is used to improve the model to deal with the imbalance in the data. For a binary prediction of flaring or non-flaring, a decision threshold is used to convert the predicted probability into a class label. The default value of 0.5 may not represent the predicted probabilities accurately due to the skewness in the data. Hence, threshold moving is used in moving the decision threshold to an optimum value, reflecting the predicted probability in the right manner. The threshold moving is done using the training data set. Figure \ref{fig:CThreshold} and \ref{fig:MThreshold} display the TSS score against threshold plots for each dataset, using which the optimum threshold has been calculated for each dataset. For the C-class dataset with varying prediction window, the threshold is 0.461 (TSS score: 0.865); for the C-class dataset with 24-hour prediction window, the threshold is 0.410 (TSS score: 0.856) and for the M-Class dataset, the threshold is 0.372 (TSS score: 0.977). Once the best threshold is achieved, the previously mentioned hyperparameters (Section \ref{subsec:lightgbm}) are tuned using the validation set to achieve the best parameters for the model.\\

\begin{figure}[hbt!]
    \centering
    \includegraphics[width=120mm]{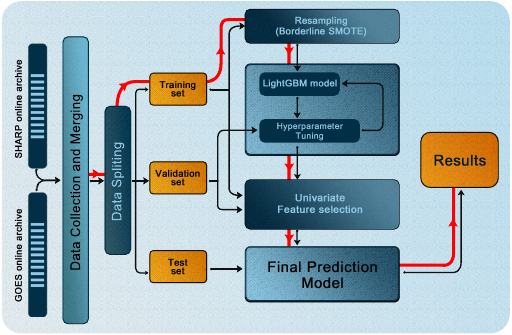}
    \caption{Flow chart of the model adopted. Starting from collection and processing solar data, the above diagram describes the steps taken throughout the paper in a brief manner.Each block represents an important step in the adopted model. The red line indicates the order in which the model was executed and the black line shows how the data was distributed between steps.}
    \label{fig:deciciontree}
\end{figure}

\begin{figure}
	\centering
	\includegraphics[width=\textwidth]{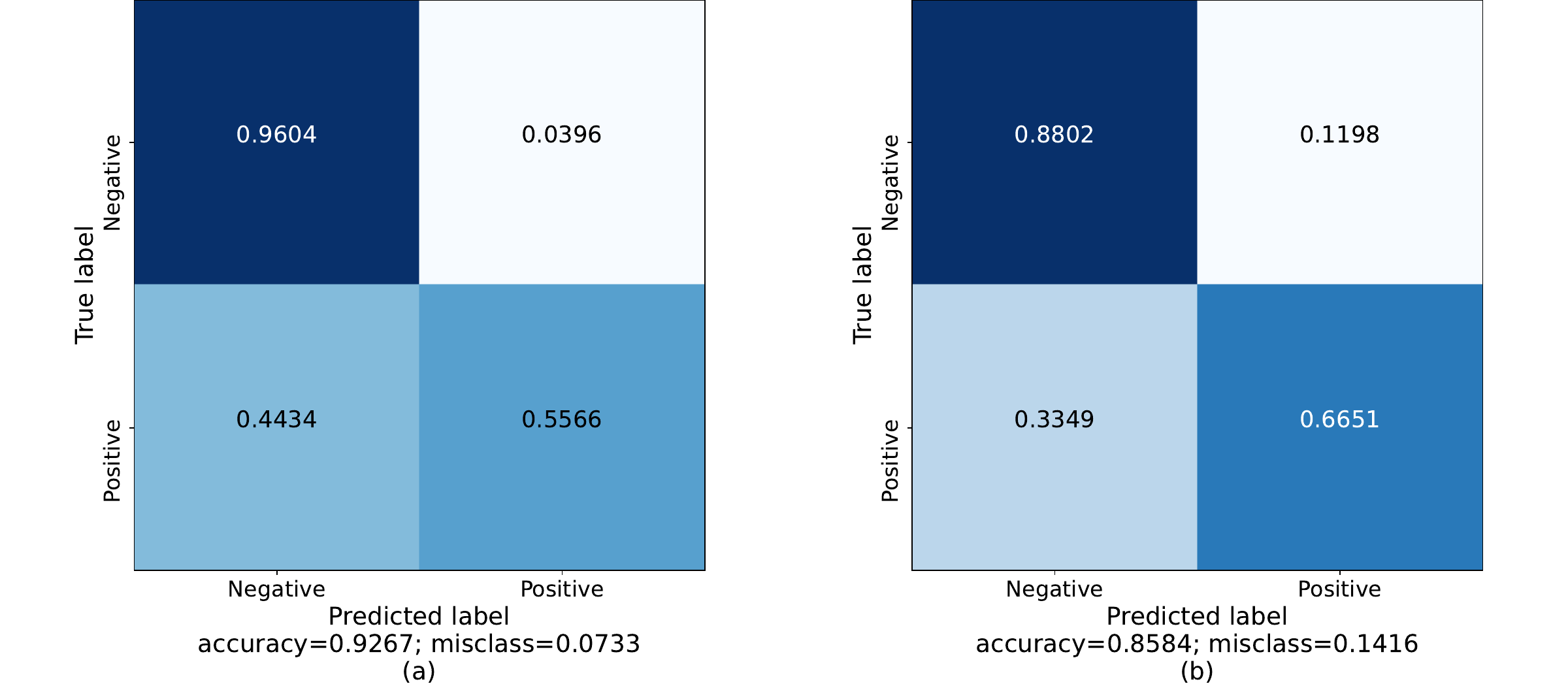}
	\caption{(a) Confusion matrix of test results using LightGBM model in the $\geq$C-Class dataset with varying prediction window; (b) Confusion matrix of test results using LightGBM model in $\geq$C-Class dataset with 24 hour prediction window.}
	\label{fig:ConfusionC}
\end{figure}

\begin{figure}[hbt!]
	\centering
	\includegraphics[width=0.45\textwidth]{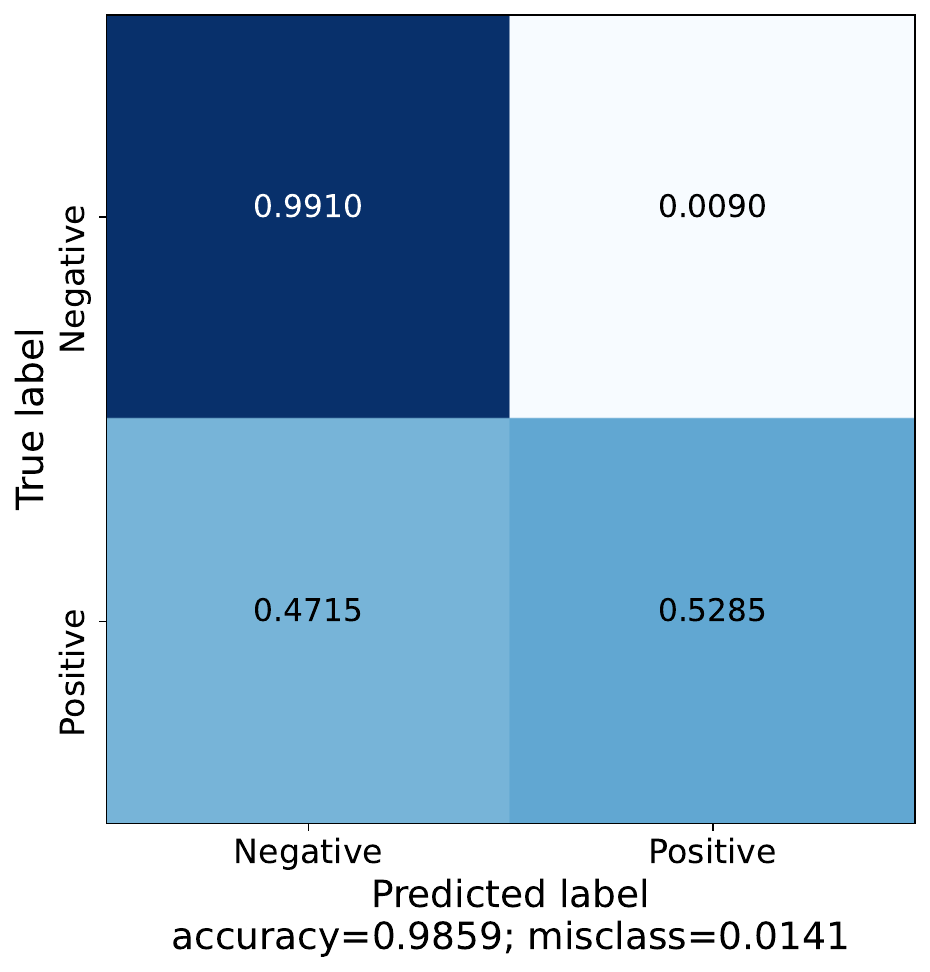}
	\caption{Confusion matrix of test results using LightGBM model in the $\geq$M-Class dataset with 24-hour prediction window.}
	\label{fig:ConfusionM}
\end{figure}

After calculating the threshold values for each dataset, we tested our model on the test set using all the parameters we calculated. All the performance metrics mentioned in the previous section were calculated to analyse the model. We calculated the score for two sets of features, only SHARP parameters and both SHARP and flaring history parameters. These scores are displayed in the Table \ref{tbl:scores}. Upon conducting a comparative analysis of the scores within both scenarios, it becomes evident that the incorporation of flaring history parameters has resulted in noteworthy enhancements across nearly all evaluated metrics, notably the TSS score. This substantiates the significance of integrating flaring history parameters and their instrumental role in improving the model performance. Along with the score, we have also provided the confusion matrix (Figure \ref{fig:ConfusionC} and Figure \ref{fig:ConfusionM}) of the classification model on each dataset. Comparing the confusion matrices between varying and 24 hour prediction windows for $\geq$C-Class, we can see that the normalised true positive value for the 24-hour prediction window surpasses that of the varying prediction window, while conversely, the normalised true negative values exhibit higher proportions for the latter approach. This observation implies that the 24-hour prediction window proves more helpful for the model in handling the class imbalance, thereby enhancing its predictive capability for positive class instances (flaring events). Conversely, in the case of the varying prediction window, despite achieving a commendable 96\% accuracy in predicting true negatives, its capacity to accurately identify positive labels stands at a relatively modest 55\%. Shifting focus to the $\geq$M-Class dataset, it becomes apparent that the model achieves an impressive 99.1\% accuracy in discerning non-flaring events, while its performance in predicting flaring events is substantially less satisfactory, amounting to a 52.8\% accuracy. This outcome can be explained by the innate rarity of M-Class events, posing challenges for the model in effectively distinguishing between positive and negative instances.  The whole workflow and the steps taken in the data analysis and model training part of this work, have been illustrated in the form of a flowchart in Figure \ref{fig:deciciontree}. In the figure, each block represents an important step in the adopted model; the black lines represent how the data is distributed in different steps, and the red line represents the order in which the data analysis steps are implemented.



\begin{center}
\begin{table}[hbt!]
	\begin{tabular}{l|p{6em}|p{7em}p{7em}p{7em}} 
		\toprule 
			    & \textbf{Metric}  & \textbf{$\geq$C-Class(varying prediction window)} & \textbf{$\geq$C-Class(24 hr prediction window)} & \textbf{$\geq$M-Class(24 hr prediction window)}  \\
		\midrule \midrule
		\multirow{7}{6em}{\textbf{SHARP and flaring history parameters}}&\textbf{Accuracy}  & 0.92421  & 0.86982  & 0.98687                      \\ 
		&\textbf{Precision} & 0.54308  & 0.41357  & 0.42171                       \\ 
		&\textbf{Recall}    & 0.57585  & 0.67212  & 0.52398                 \\ 
		&\textbf{ROC AUC}   & 0.88703  & 0.88053  & 0.92618                       \\ 
		&\textbf{TSS}       & 0.53173  & 0.56429  & 0.51597                       \\ 
		&\textbf{BACC}      & 0.76586  & 0.78214  & 0.75799                       \\ 
		&\textbf{HSS}       & 0.51731  & 0.44170  & 0.46012                 \\ \midrule
		\multirow{7}{6em}{\textbf{SHARP parameters}}&\textbf{Accuracy}  & 0.91440  & 0.85773  & 0.97987                       \\ 
		&\textbf{Precision} & 0.48841  & 0.38504  & 0.26559                       \\ 
		&\textbf{Recall}    & 0.56554  & 0.67062  & 0.47561                       \\ 
		&\textbf{ROC AUC}   & 0.86081  & 0.86477  & 0.92951                      \\ 
		&\textbf{TSS}       & 0.51165  & 0.54949  & 0.46107                       \\ 
		&\textbf{BACC}      & 0.75583  & 0.77474  & 0.73053                       \\ 
		&\textbf{HSS}       & 0.47729  & 0.41345  & 0.33084                       \\ 
		\bottomrule
	\end{tabular}
	\caption{Flare prediction results of LightGBM model on different datasets}
	\label{tbl:scores}
\end{table}
\end{center}
\subsection{Feature set evaluation}
\begin{figure}[hbt!]
    \centering
    \includegraphics[width=\textwidth]{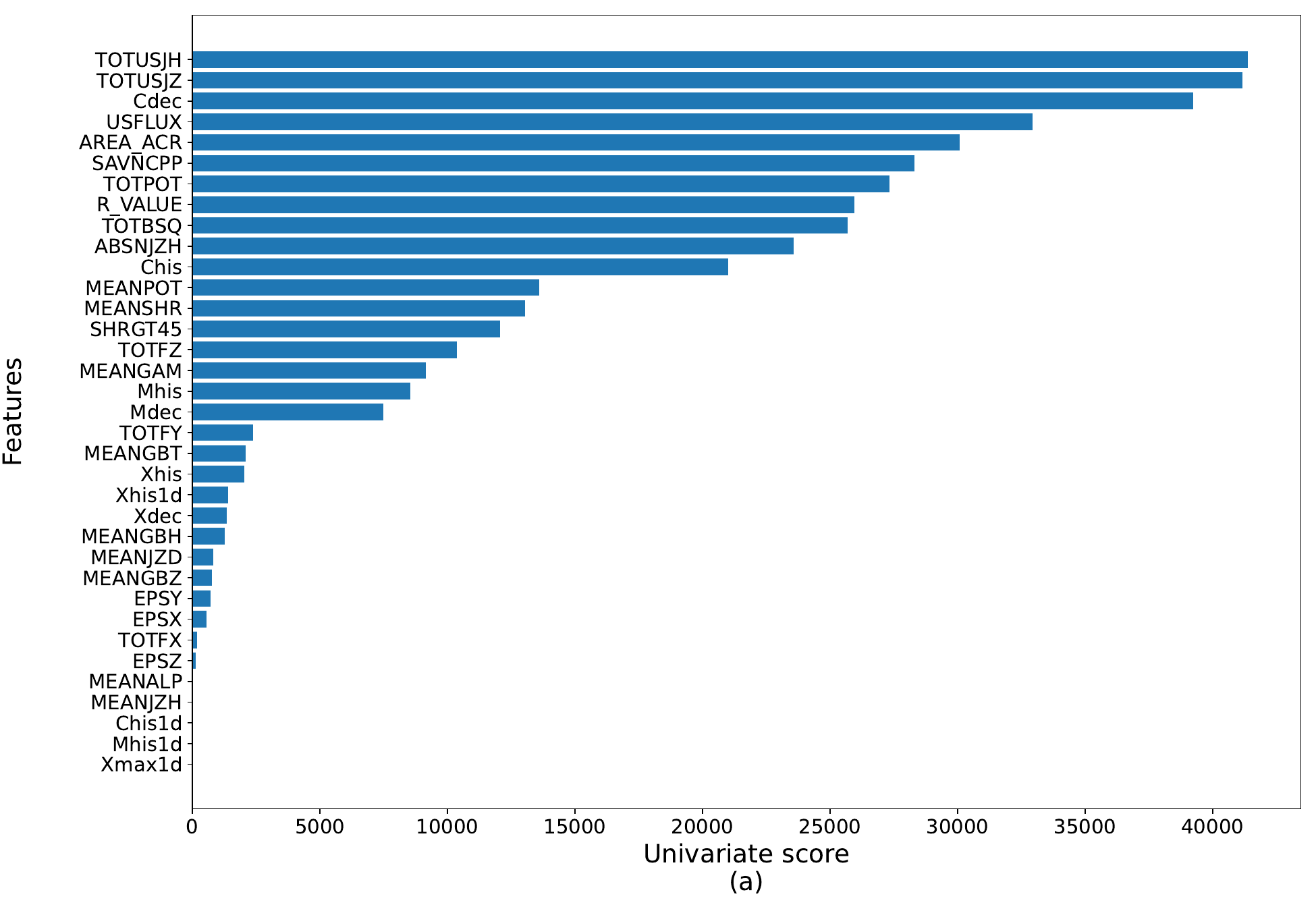}
    \caption{Univariate feature importance graph for predicting $\geq$C-class with 24 hour prediction window. The bar lengths indicate individual ANOVA F-value score.}
    \label{fig:featuresCfull24}
\end{figure}

Not all features are important towards solar flare prediction \citep{bobra_2015_solar}. While some of them show high correlation with the output, for a few of the features, the inclusion of them in the feature could decrease the overall performance. Thus we use a univariate feature selection algorithm, using ANOVA F-value score to rank them. For $\geq$C-Class (24-hour prediction window) dataset (Figure \ref{fig:featuresCfull24}), the Total unsigned current helicity feature is identified as having the highest F-score (TOTUSJH). Numerous studies have suggested a strong correlation between the accumulation of magnetic helicity and the occurrence of flares in active regions \citep{Park_2010, Liu_2023}. Features like Total magnetic vertical current (TOTUSJZ), Total unsigned flux (USFLUX) are also among the high-ranked features for this dataset. From the flaring history parameters, CDEC was ranked in the top 3, suggesting the time decay value of C flares in an active region shows high correlation towards a flaring event. It can also be seen that the history of C-flares is also ranked high.  It can be noted that just considering the SHARP active region parameters, 11 out of the top 13 magnetic summary features selected matched with the top 13 magnetic summary features mentioned in \cite{Bobra_2014}. These include TOTUSJH, TOTUSJZ, USFLUX, AREA\_ACR, SAVNCPP, TOTPOT, R\_VALUE TOTBSQ, ABSNJZH, MEANPOT, and MEANSHR. Thus our findings are consistent with published values in the literature. \\

\begin{figure}[hbt!]
    \centering
    \includegraphics[width=\textwidth]{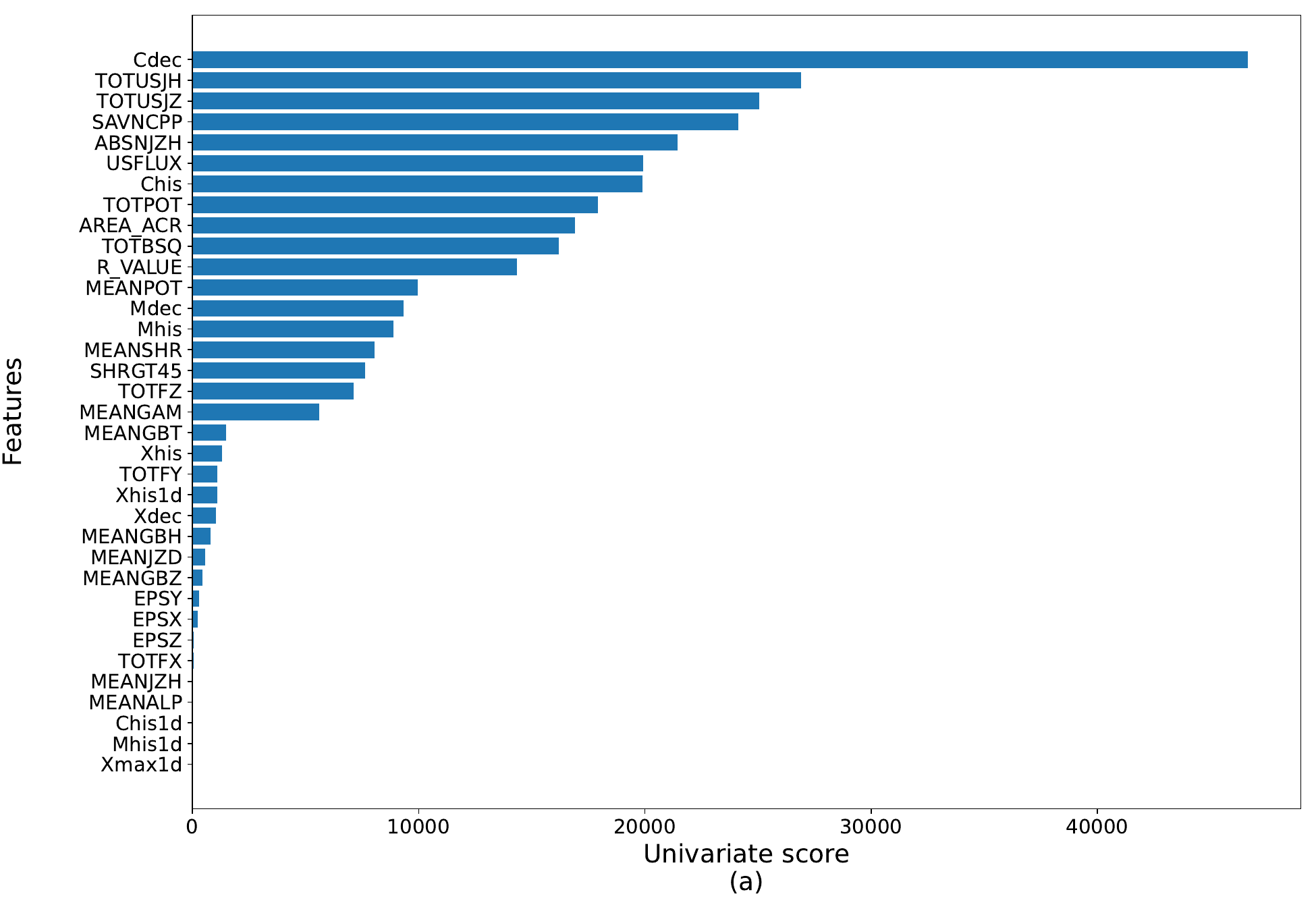}
    \caption{Univariate feature importance graph for predicting $\geq$C-class with varying prediction window. The bar lengths indicate individual ANOVA F-value score.}
    \label{fig:featuresCfullcurve}
\end{figure}

For $\geq$C-Class (varying prediction window) dataset (Figure \ref{fig:featuresCfullcurve}) an interesting trend is observed. Even though they display slightly different importance, all the first 12 features are the same as the first 12  features in the previous dataset. This shows that the basic dependence of different magnetic and flaring history features on flaring activity remains intact with the change in the prediction window. But a notable difference occurs in the importance of the CDEC parameter. Compared to the 24-hour prediction window, the varying prediction window shows higher relative importance towards the C flare decay value. This could have derived from the fact that, during high solar activity, the 24-hour prediction window could have included a higher number of samples even though they are from different C Class flares, thus reducing the importance of CDEC parameter. By the introduction of varying prediction windows at higher solar activity, only the samples corresponding to observed flare will contribute, thus letting the model learn it as an important feature for flare prediction.\\

\begin{figure}[hbt]
    \centering
    \includegraphics[width=\textwidth]{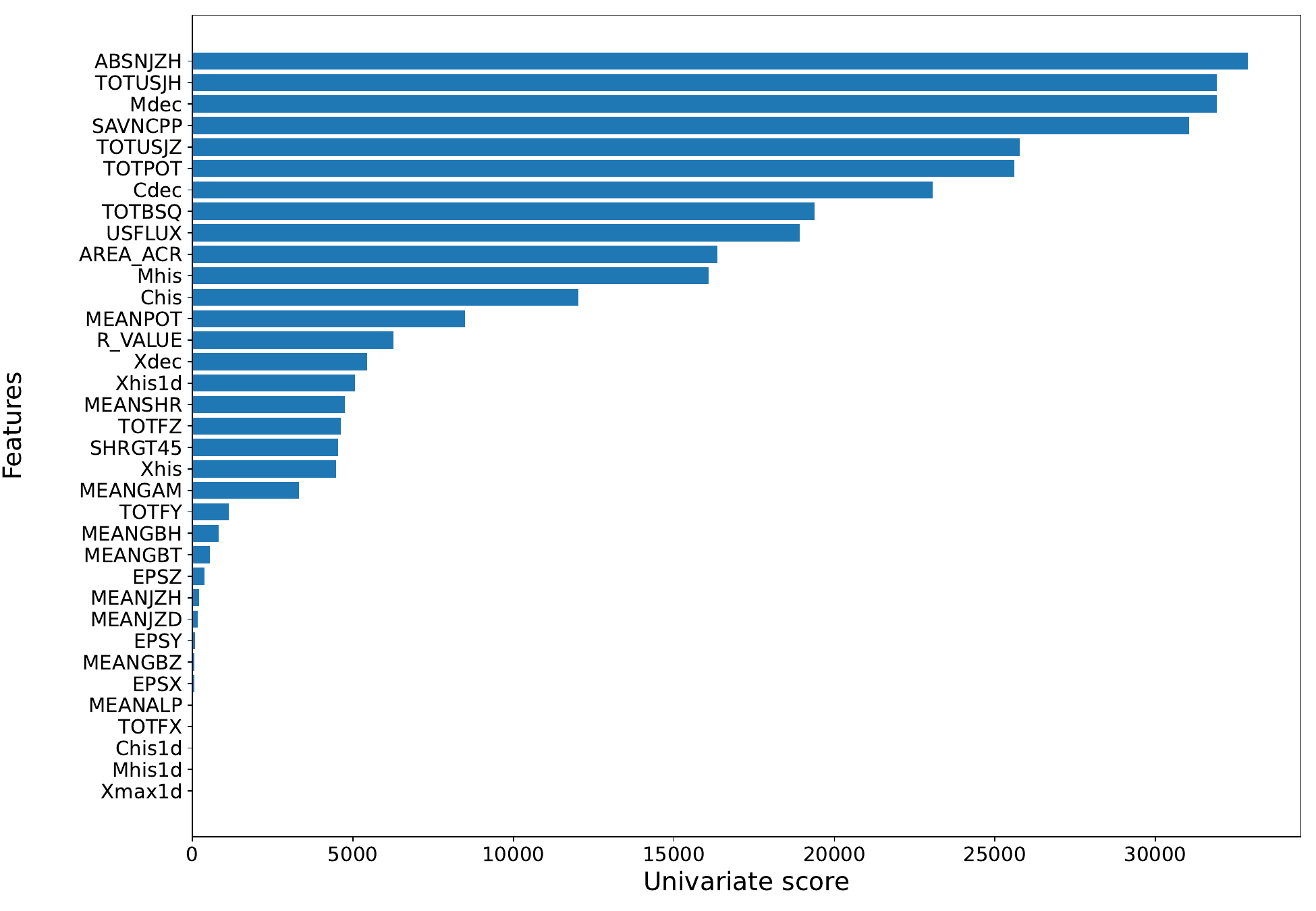}
    \caption{Univariate feature importance graph for predicting $\geq$M-class with 24 hour prediction window. The bar lengths indicate individual ANOVA F-value score.}
    \label{fig:featuresMfull}
\end{figure}

For $\geq$M-Class (24 hour prediction window) dataset (Figure \ref{fig:featuresMfull}), the feature with the highest F-score is Absolute value of the net current helicity (ABSNJZH). Features like Total unsigned current helicity (TOTUSJH), Total magnetic vertical current (TOTUSJZ), and Sum of the modulus of the net current per polarity (SAVNNCPP) also are among highly ranked features for this dataset. From the flaring history parameters, similar to C-Class prediction,  time decay value of M flares in an active region (MDEC) shows high correlation towards an M Class flaring event. We also observe that apart from M Class flaring history parameters, CDEC and CHIS also show a higher importance towards the M Class flaring events. This could imply that an active region with C class flares occurring frequently is more prone to M class events.\\

\begin{figure}[hbt!]
    \centering 
    \includegraphics[width=\textwidth]{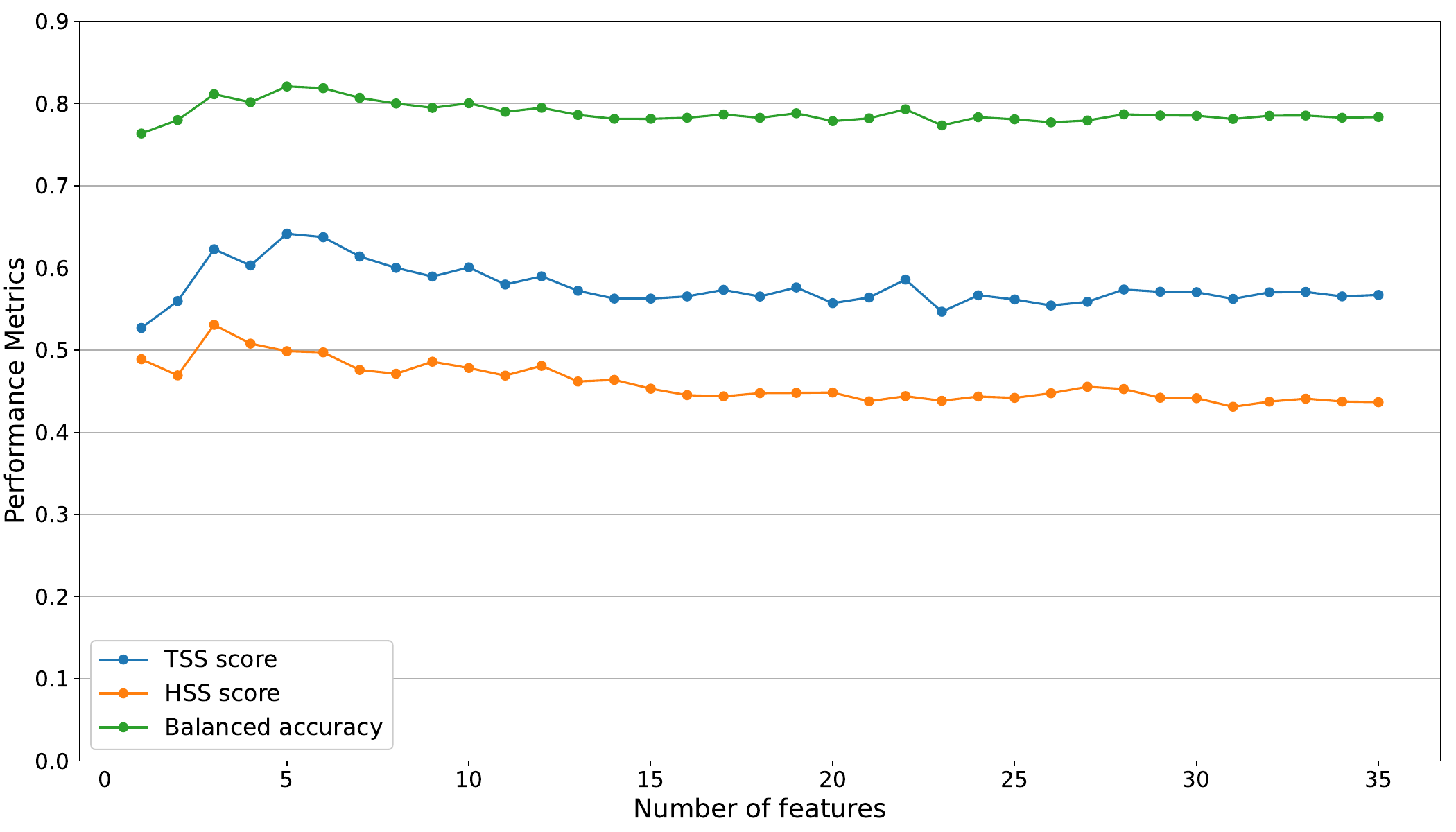}
    \caption{Performance metrics (TSS, HSS and BACC) as a function of number of features using LightGBM classifier for $\geq$C-Class with 24 hour prediction window dataset.}
    \label{fig:Cfeatscore24}
\end{figure}

\begin{figure}[hbt!]
    \centering
    \includegraphics[width=\textwidth]{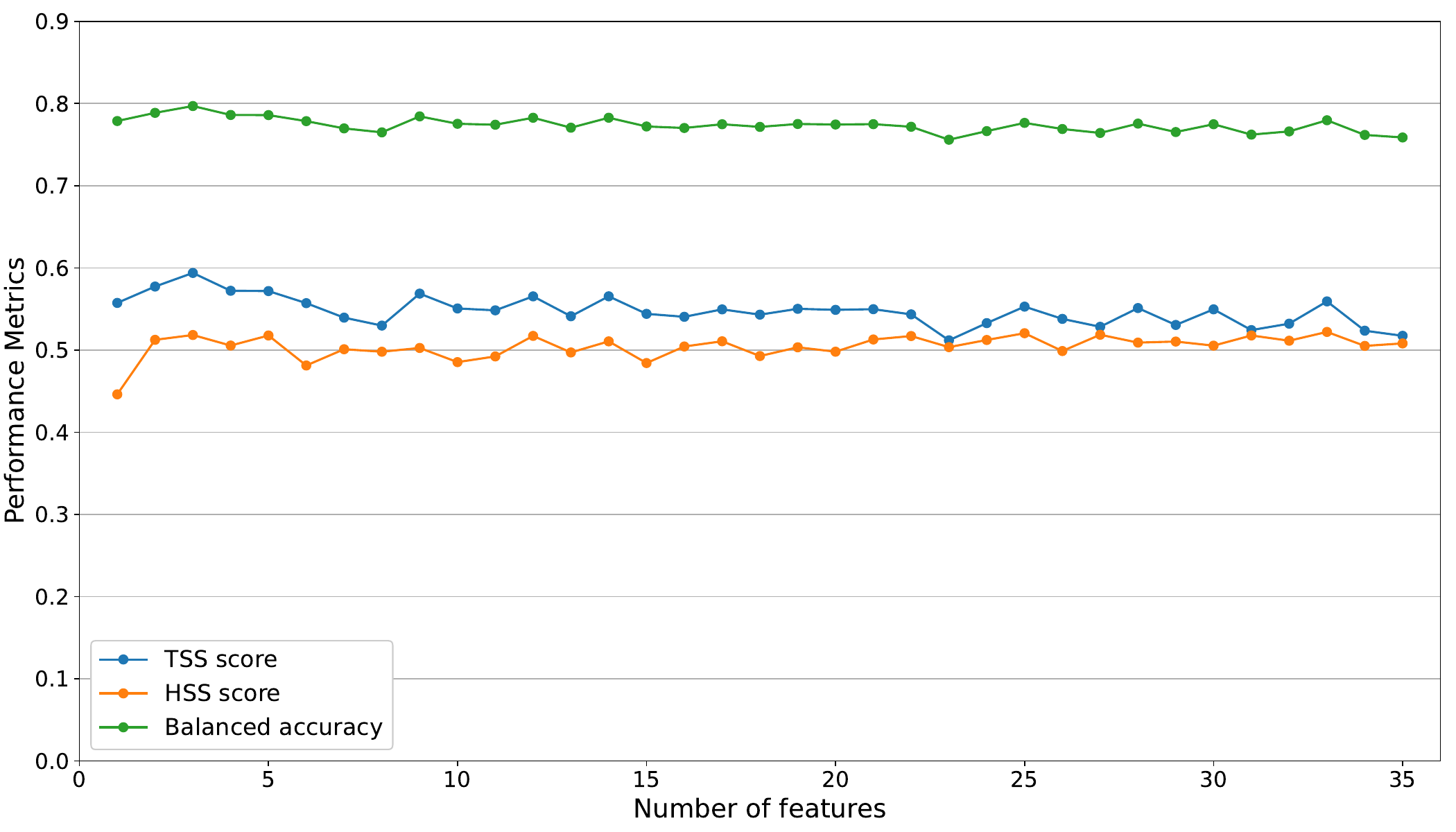}
    \caption{Performance metrics (TSS, HSS and BACC) as a function of number of features using LightGBM classifier for $\geq$C-Class with varying prediction window dataset.}
    \label{fig:Cfeatscore_curve}
\end{figure}

\begin{figure}[hbt!]
    \centering
    \includegraphics[width=\textwidth]{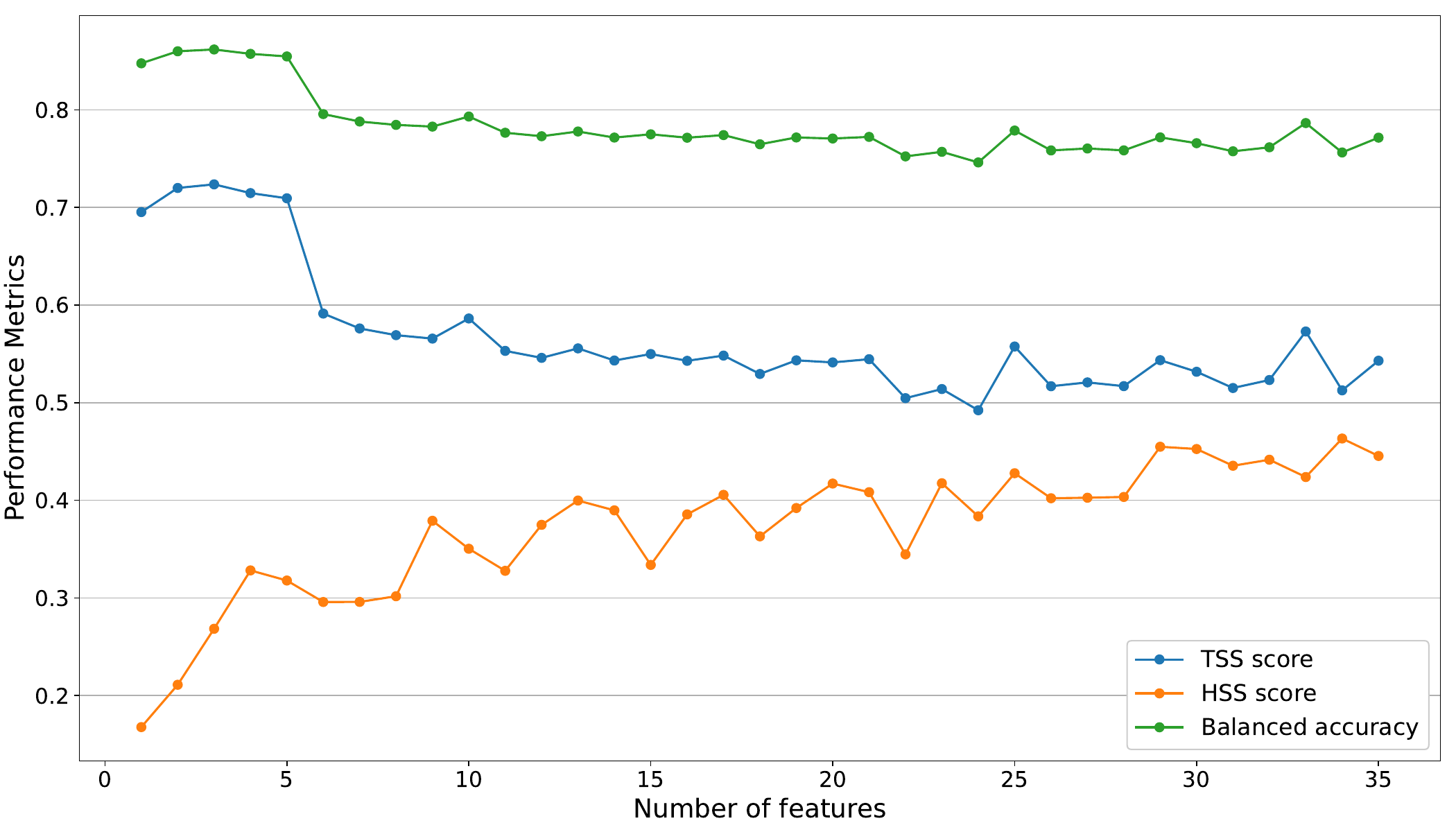}
    \caption{Performance metrics (TSS, HSS and BACC) as a function of number of features using LightGBM classifier for $\geq$M-Class with 24 prediction window dataset.}
    \label{fig:Mfeatscore24}
\end{figure}

In order to identify the optimal feature set, we have conducted an evaluation of performance metrics that are contingent upon the number of features utilised. This analysis entails arranging features in descending order with regard to their significance. The result can be seen in Figure \ref{fig:Cfeatscore24}, \ref{fig:Cfeatscore_curve} and \ref{fig:Mfeatscore24}. This brings up an interesting property that not all the parameters are required to produce the best classifier. In the context of the $\geq$C-Class dataset, encompassing both variable and 24-hour prediction windows, noticeable trends emerge regarding feature selection. Initially, an ascent in scores is observed for 1 to 5 features, followed by a subsequent plateauing effect as additional features are incorporated. This phenomenon potentially underscores the pivotal contribution of the initial high-importance features towards predicting solar flaring activity. Conversely, the inclusion of less significant features appears to lead to a saturation of scores, indicating their comparatively diminished role in flare prediction. In the case of the $\geq$M-class dataset, a divergence in trends is evident across different performance metrics. Specifically, BACC and TSS, designed to address imbalanced datasets, exhibit a parallel trajectory, with scores peaking for 1 to 5 features and gradually diminishing thereafter. This trend implies that incorporating features beyond the most crucial ones not only stabilises but also reduces the predictive scores. This decline may stem from potential overfitting, as the model's complexity increases, leading to the incorporation of less relevant patterns from the less important features. While both the $\geq$C-Class and $\geq$M-Class datasets have undergone resampling to achieve comparable ratios, the relative strength of this trend appears to be less pronounced in the former. This discrepancy may be attributed to the specific manner in which SMOTE algorithms perform the resampling process. SMOTE primarily operates by generating synthetic data points through interpolation between existing data vectors. Although this technique mitigates class imbalance concerns, it does not fundamentally alter the underlying distribution of parameters within the dataset. Given that events falling into the $\geq$M-Class category are inherently less frequent than those categorised as $\geq$C-Class, this results in a lower coverage of parameter space for the former class. Consequently, the model may encounter challenges in discerning and learning relevant patterns, particularly from the less prominent features within the dataset. In contrast to the other two metrics, the HSS demonstrates a distinct trajectory, implying varying influences of features and complexity on this metric itself.\\

\clearpage

\

\begin{figure}[hbt!]
    \centering
    \includegraphics[width=\textwidth]{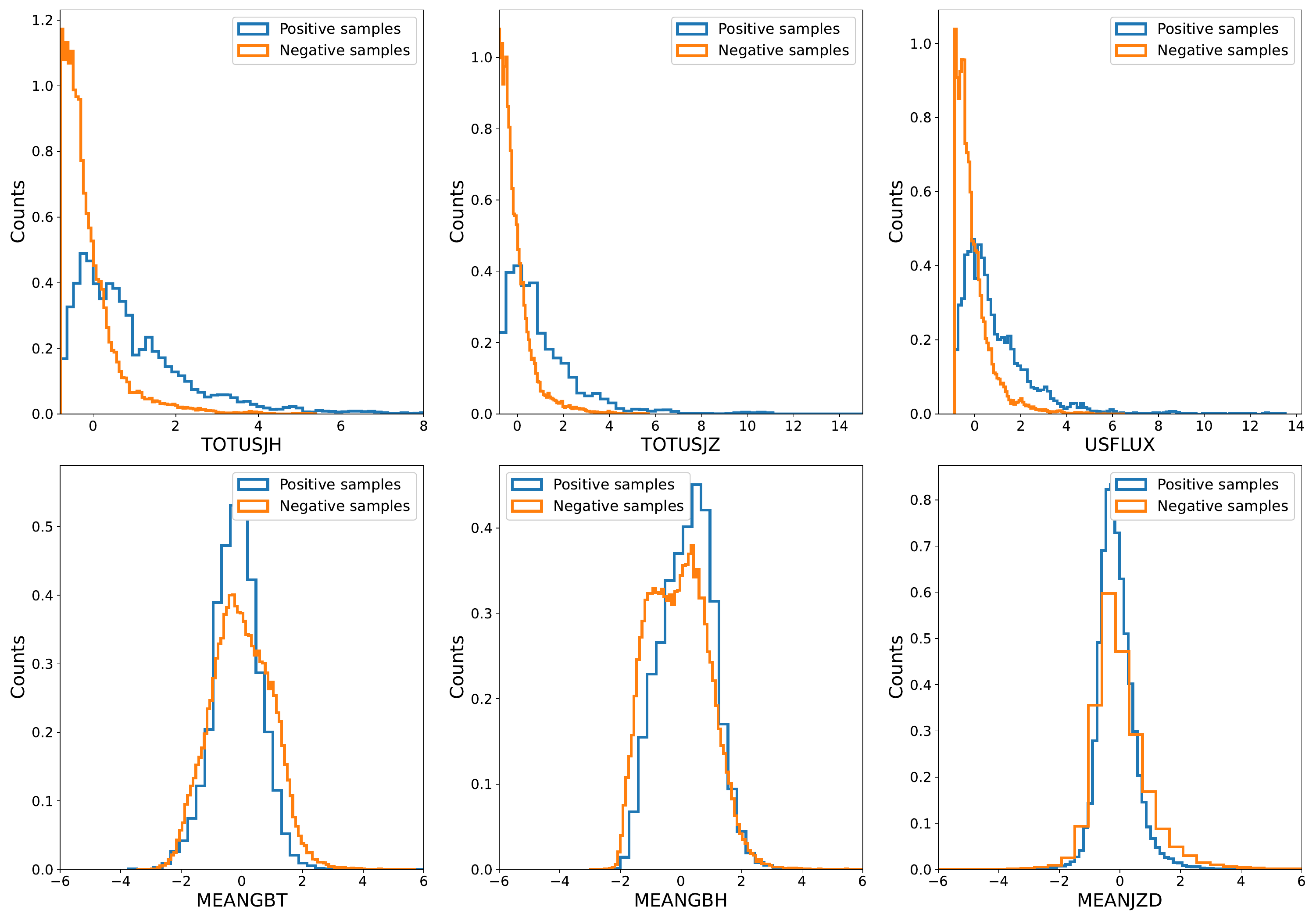}
    \caption{Histogram plot of various features in both positive and negative sample in dataset 1 ($\geq$C-Class with 24 hour varying window). The top row contains the parameters from the selected set of parameters while the bottom row contains the rejected parameters.}
    \label{fig:Chistogram}
\end{figure}
\begin{figure}[hbt!]
    \centering
    \includegraphics[width=\textwidth]{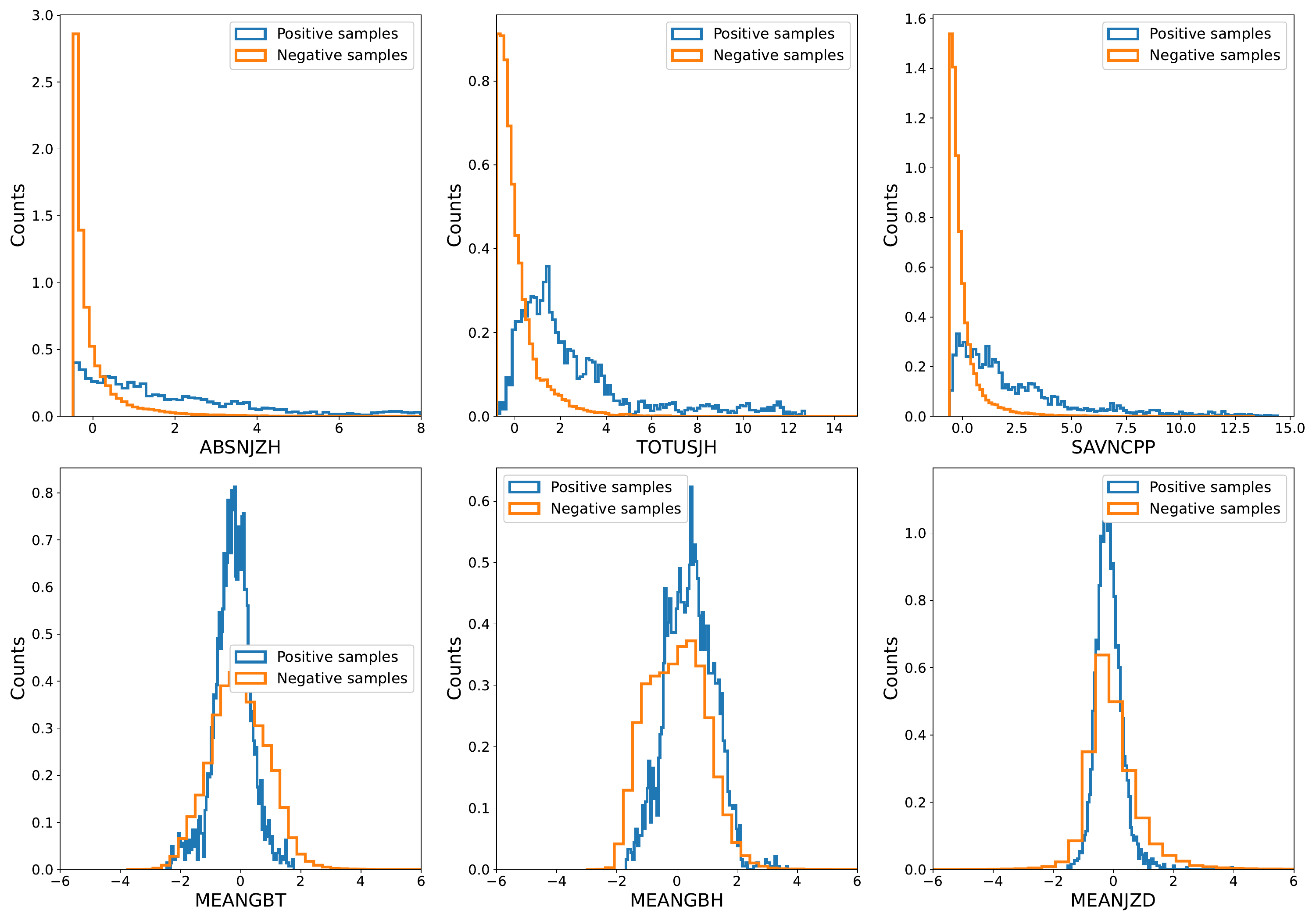}
    \caption{Histogram plot of various features in both positive and negative sample in dataset 2 ($\geq$M-Class with 24 hour varying window). The top row contains the parameters from the selected set of parameters while the bottom row contains the rejected parameters.}
    \label{fig:Mhistogram}
\end{figure}

In an attempt to elucidate the detectable dissimilarity in the importance of various features, we have undertaken the visualisation of feature importance distributions concerning both positive (flaring) and negative (non-flaring) labels. This visual representation is presented in Figures \ref{fig:Chistogram} and \ref{fig:Mhistogram}. Notably, for features attributed with heightened importance, a noticeable divergence in distribution peaks and spreads between the positive and negative labels is observed, thereby conferring distinctiveness to these features. Conversely, among features deemed less significant, a marked semblance in distribution patterns is discernible, indicative of their diminished discriminative capacity. This disparity in feature importance manifestation is particularly pronounced within the context of the M-Class dataset, wherein the broader distribution spread of certain features is evident. This phenomenon is particularly notable due to the scarcity and heightened magnitude of $\geq$M-class flares in contrast to $\geq$C-Class flares.

\begin{center}
\begin{table}[hbt!]
	\begin{tabular}{p{7em}|p{9em}p{9em}p{9em}} 
		\toprule 
		       \textbf{Metric}  & \textbf{$\geq$C-Class(varying prediction window)} & \textbf{$\geq$C-Class(24 hr prediction window)} & \textbf{$\geq$M-Class(24 hr prediction window)} \\
		\midrule \midrule
		\textbf{Accuracy}  & 0.92055  & 0.89841  & 0.96987                    \\ 
		\textbf{Precision} & 0.52083  & 0.50001  & 0.22541                    \\ 
		\textbf{Recall}    & 0.58531  & 0.63108  & 0.71869                      \\ 
		\textbf{ROC AUC}   & 0.89077  & 0.88711  & 0.94186                    \\ 
		\textbf{TSS}       & 0.53632  & 0.63057  & 0.69134                        \\ 
		\textbf{BACC}      & 0.76816  & 0.80979  & 0.84567                        \\ 
		\textbf{HSS}       & 0.50769  & 0.45664  & 0.33201                        \\ \midrule
		\bottomrule
	\end{tabular}
	\caption{Flare prediction results of LightGBM model using the highest scored parameters}
	\label{tbl:scores2}
\end{table}
\end{center}
\begin{figure}
	\centering
	\includegraphics[width=\textwidth]{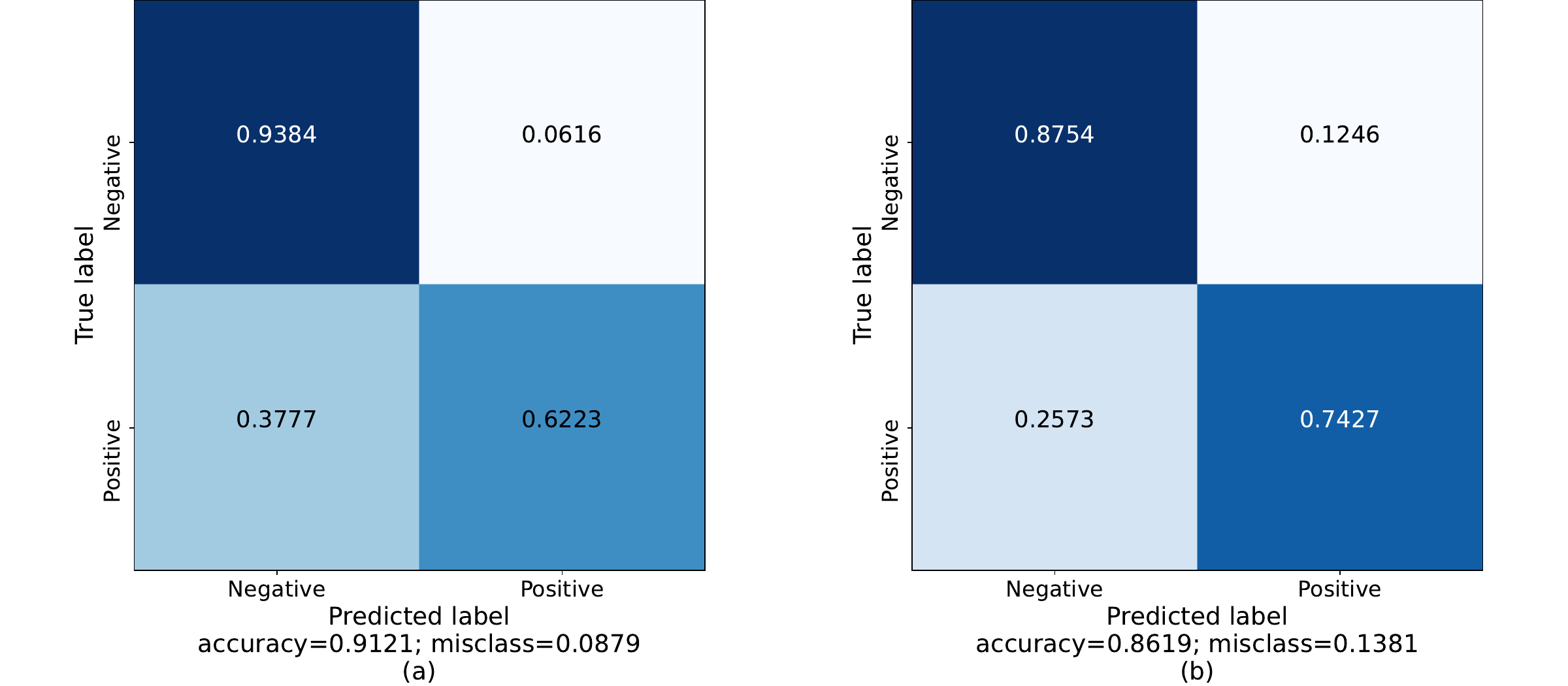}
	\caption{(a) Confusion matrix of test results using the highest scoring parameters in $\geq$C-Class dataset with varying prediction window; (b) Confusion matrix of test results using the highest scoring parameters in $\geq$C-Class dataset with 24 hour prediction window.}
	\label{fig:ConfusionCselected}
\end{figure}
\begin{figure}[hbt!]
	\centering
	\includegraphics[width=0.45\textwidth]{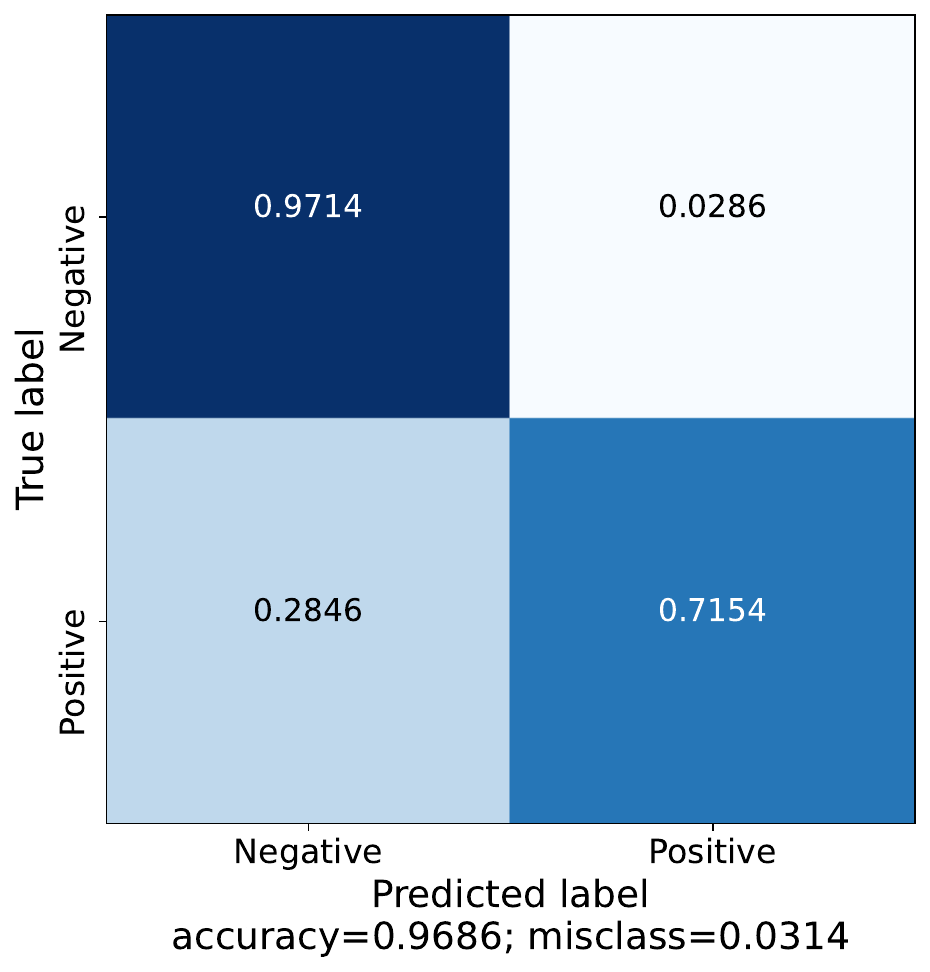}
	\caption{Confusion matrix of test results using the highest scoring parameters in $\geq$M-Class dataset with 24-hour prediction window.}
	\label{fig:ConfusionMselected}
\end{figure}

After the above analysis, we selected the best set of features as per data in Figures  \ref{fig:Cfeatscore24}, \ref{fig:Cfeatscore_curve}, and \ref{fig:Mfeatscore24}. Based on the plots, we identified the optimal number of features around the highest TSS metric within each dataset. From the figures, it is clear that all three datasets displayed the highest TSS metric in between 3 - 5 sets of features. To have a more robust and general model and avoid having too few features leading to loss of information, a set of top five features was selected for each dataset.Subsequently, the classifier model was re-trained exclusively using these selected prominent features. The subsequent outcomes obtained utilising this refined feature subset are documented in Table \ref{tbl:scores2}. We have also plotted the confusion matrices (Figure \ref{fig:ConfusionCselected} and \ref{fig:ConfusionMselected}) for the new subset of features for comparison. When comparing with Table \ref{tbl:scores}, concerning the $\geq$C-Class dataset, consistent enhancements in pivotal metrics, namely, TSS and BACC, are evident for both labeling algorithms. This observation underscores the efficacy of utilising a constrained feature subset, whereby improved predictive power is achieved in contrast to employing the complete set of features. Regarding the remaining metrics, precision, gauging the proportion of true positives among positive predictions, demonstrates better scores in the recent model for both scenarios. This outcome indicates the model's heightened success in accurately identifying positive instances. In terms of recall, which quantifies the fraction of correctly identified actual positives, a minor reduction is discernible in the recent model due to a slight elevation in False Negative values. As for the HSS metric, a marginal increase is noted in the 24-hour prediction window dataset, whereas a slight decrease is observed in the varying window dataset. Given HSS's susceptibility to data imbalance, these trends inadequately represent model performance.  
Analysing the confusion matrices (Figure \ref{fig:ConfusionCselected}) in relation to its predecessor (Figure \ref{fig:ConfusionC}), it becomes evident that while a marginal decline in the accuracy of True Negatives is noticeable, the model exhibits a significantly enhanced capability to predict True Positives within both the varying prediction window and 24-hour prediction window datasets. \\

With regard to the $\geq$M-Class dataset, a notable finding is the significant increments in TSS and BACC metrics, which point to a noticeable improvement in model performance. Conversely, the metric of precision experiences a significant reduction, which can be attributed to the interplay between the scarcity of $\geq$M flares and a slight decrement in true negative values. This compound effect contributes to the observed decline in precision. Similar mechanisms are also evident in the context of recall, where a marked increase is witnessed for the recent model. This phenomenon can be attributed to analogous underlying factors. The trend holds true for the Heidke Skill Score (HSS) metric as well, exhibiting a declining trajectory. It is imperative to emphasise that, during performance evaluation, metrics other than TSS and BACC should not be utilised due to their unsuited handling of data imbalance. Upon inspecting the confusion matrix, it becomes evident that the predictive capacity has notably improved, underscored by a significant enhancement in true positives. 

\subsection{Comparison with other similar works}

We have also conducted a comparison of our results using the LightGBM model with existing studies that employed different methods, such as the Support Vector Machine (SVM) algorithm, Random Forest (RF), Long-Short Term Memory (LSTM), k-NN method, and Deep Flare Net (DeFN). Various studies have used diverse sets of scores or metrics to evaluate their findings. Given our aim to align our outcomes with related research, we have adopted the most commonly used evaluation measures: True Statistical Score (TSS) and Accuracy (Acc) for comparative analysis.\\

It is important to note that different studies in the literature have constructed datasets in varying ways, involving distinct time intervals for observations and analysis periods. Therefore, we have considered the most favourable outcomes achieved by these studies on their respective testing datasets for events of magnitude $\geq$M within a 24-hour period. While the presented analysis may not provide an exhaustive direct comparison of different solar flare forecasting models due to variations in data construction and analysis, our effort is to offer insights based on the chosen performance measures.\\

Among the available models, the Support Vector Machine (SVM) algorithm has been widely used. \cite{bobra_2015_solar} used a method to select important factors and found that using 13 of these features, they got a TSS of 0.761 and Acc of 0.924. \cite{nishizuka_2017_solar} also used the SVM algorithm, mixing vector magnetogram and UV brightening data to get a TSS of 0.87 and Acc of 0.988. Another study by \cite{florios_2018_forecasting} used SVM too, and they used data over five years with 3-hour gaps to forecast $\geq$M flares, obtaining a TSS of 0.59 and Acc of 0.94. Additionally, \cite{ribeiro_2021_machine} reported a TSS of 0.622 using SVM.\\

Some studies tried the Long Short-Term Memory (LSTM) network to predict whether an Active Region (AR) will produce a $\geq$M class flare within 24 hours. \cite{liu_2019_predicting} used this method, including 25 SHARP parameters and 15 flare history parameters, to achieve a TSS of 0.79 and Acc of 0.909.\\

The Random Forest (RF) algorithm has also been used for creating solar flare prediction models using machine learning. \cite{florios_2018_forecasting} used RF with SDO/HMI SHARP data, getting a TSS of 0.74 and Acc of 0.93. Furthermore, \cite{ribeiro_2021_machine} achieved a TSS of 0.63 for $\geq$M class solar flare prediction within 24 hours using RF.\\

Alongside SVM, \cite{nishizuka_2017_solar} explored k-Nearest Neighbors (k-NN) with a TSS of 0.912 and Acc of 0.995, and Extremely Randomised Trees (ERT) with a TSS of 0.71 and Acc of 0.990. Using 65 features, they found that k-NN performed better than ERT and SVM. Additionally, \cite{nishizuka_2018_deep} used extra features like hot coronal brightening and data from SDO/AIA 131 $\text{\normalfont{\AA}}$ and GOES X-ray emissions, to create the Deep Flare Net (DeFN) algorithm. This approach achieved a TSS of 0.802 and Acc of 0.858.\\

In our study, we used LightGBM and got a TSS of 0.69 and Acc of 0.970 by focusing on the top features. Similarly, \cite{ribeiro_2021_machine} used LightGBM and got a TSS of 0.61. They also found that SVM, RF, and LightGBM gave similar TSS results.\\

To provide a complete picture of solar flare predictions using different methods, we used a modified Taylor diagram \citep{taylor2001summarizing}. This kind of diagram is often used to compare how well a model's forecasts match real observations.  Usually, the Pearson correlation coefficient (cc) and standard deviation (std) between model’s result and observation is considered. In this study, we have taken a similar approach but have used Acc and TSS values instead of cc and std.\\

\begin{figure}[hbt!]
	\centering
	\includegraphics[width=0.7\textwidth]{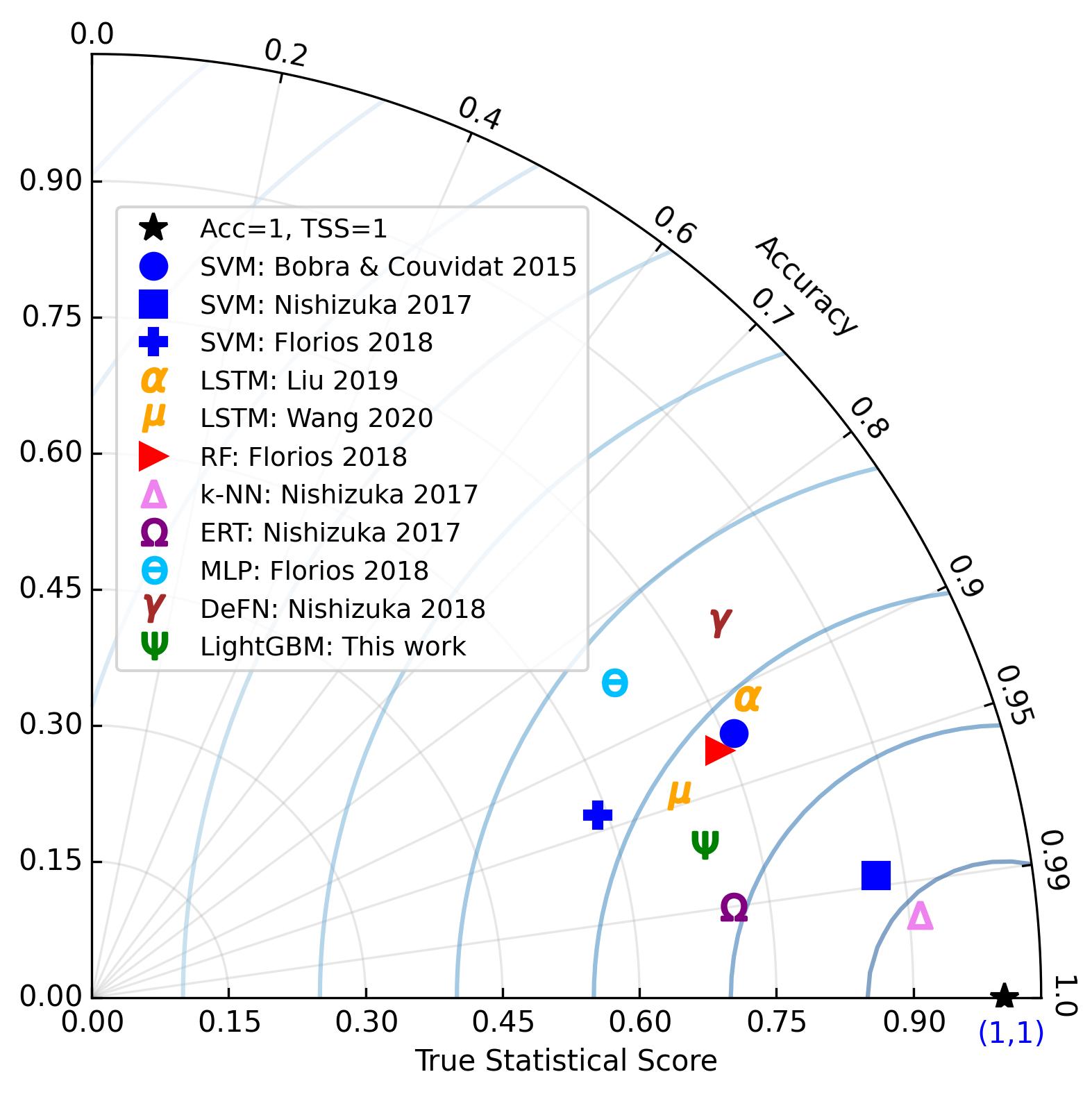}
	\caption{The picture depicts a modified Taylor diagram for comparing the results of solar flare prediction models using different machine learning algorithms. The radial distances from the center represent their TSS values and azimuthal positions are corresponding to Acc of their models. The position where TSS=1 and Acc=1 has been taken as the reference point for comparison.}
	\label{fig:Taylorplot}
\end{figure}

Figure \ref{fig:Taylorplot} illustrates a modified Taylor diagram presenting outcomes from distinct algorithms in comparable studies. The reference point, denoted by a black star, is set at TSS=1 and Acc=1. The distance of the marks from the center reflects their TSS value, while their position around the circle signifies their Acc value. The circles centered on the reference point indicate how closely the model outcomes resemble the ideal state of Acc=1 and TSS=1. The figure shows the results from those works in which both TSS and Acc were mentioned.\\

Models that are closer to the reference point, or essentially within smaller circles, exhibit stronger resemblance to Acc=1 and TSS=1. This can be a method to compare diverse model performances based on their best results. Interestingly, among the three works utilising SVM, they present notably different results, each lying in distinct circles. Conversely, studies using LSTM demonstrate comparable outcomes, all positioned within the same circle at almost identical distances from the reference point.\\

Following this approach, our LightGBM-based model outperforms many prior studies, especially those utilising LSTM, RF, MLP, and DeFN. Notably, the results of \cite{nishizuka_2017_solar} study showcase the highest performance. In this study, they employed k-NN, SVM, and ERT, with k-NN delivering the most favourable results.\\

The strong performance of the algorithms in \cite{nishizuka_2017_solar} study could be attributed to their use of a wide range of solar features. They employed a total of 65 features and demonstrated their relative importance for solar flare prediction. Their findings revealed that past flare activities, like the flare history in an Active Region (AR) and the highest X-ray intensity of the previous day, held the greatest significance. Following this, the configurations of magnetic neutral lines, unsigned magnetic flux, and the area of UV brightening were also considered highly important.\\

While their findings pointed to the greater significance of certain features compared to others, their results also indicated that employing all features might improve performance. In contrast, other studies have shown that utilising a more focused set of essential features could lead to equal or even better outcomes. For instance, \cite{bobra_2015_solar} explored a wide range of 25 parameters and found that using only four—total unsigned current helicity, total magnitude of the Lorentz force, total photospheric magnetic free energy density, and total unsigned vertical current—resulted in a TSS score similar to the combined top 13. Additionally, \cite{liu_2019_predicting} demonstrated that using the 14-22 most important features, including both flare history and magnetic parameters, yielded superior performance compared to using all 40 features together (25 SHARP and 15 flare history). Our results also indicate that excluding less valuable features from the set could enhance the model's performance in predicting solar flares.\\

\section{Summary and Conclusion}\label{sec:SnC}
In this work, we used a gradient boosted decision tree classifier known as LightGBM to develop a machine learning model to predict solar flares based on observed magnetic and flaring history parameters. Using the data from SDO HMI SHARP data archive and GOES solar flare database, we prepared the final datasets for the machine learning model. Along with the 25 magnetic parameters, we also calculated 12 derived parameters to also account for the flaring history of that particular active region. Solar flare prediction is the exact definition of an imbalanced classification problem in machine learning. This arises due to the fact that the number of non-flaring events or weak flare events (A and B class flares) far exceed the strong flares. This introduces an innate bias towards non flaring/weak flaring events while training a machine learning model. We used various techniques in preprocessing as well as in the classifier to address this issue. Borderline Synthetic Minority Over-sampling Technique (Borderline SMOTE) along with random undersampling was employed as a preprocessing step to get the ratio of the flaring and non-flaring events to be around 0.6. This reflected as an increase in the model performance thus was accepted as a part of the final classification pipeline.\\

For labeling the data as positive and negative samples, along with the commonly used operational algorithm we experimented with a varying labeling system which takes into account the varying levels of solar activity. Thus the model was trained and tested on three datasets: $\geq$C-class(varying prediction window), $\geq$C-Class (24 hour prediction window) and M-Class (24 hour prediction window). In the varying prediction window algorithm, the prediction window varies from 24 hours to 6 hours and then back to 24 hours while moving from the start to the end of solar cycle. This attains a minimum of 6 hours prediction window at solar maximum. While using the classifier on both the labeling methods with $\geq$C-Class dataset, the scores display an interesting trend where both 24 hour prediction window and the varying prediction window produce TSS scores very close to each other. This could be a result of the fact that as flares are more frequent during activity peak time, due to faster changes in magnetic activity, a 6 hr prediction window is sufficient in place of a 24 hour prediction window. But the varying prediction algorithm underachieves in the case of $\geq$M-Class, which can point to fact that, being a stronger class of flare, a constant 24 hour window is necessary to capture the changes in the magnetic properties as opposed to $\geq$C-class. It should be noted that the present study does not extensively elaborate upon this case, as the utilisation of the varying prediction algorithm yielded notably low TSS scores, consequently rendering the algorithm inconsequential within the context of the examined scenario..\\

In addition to classification, we also performed feature importance study for all the magnetic and flaring history parameters. We used ANOVA F-value score to rank the features for all three datasets. Majority of the most important features aligned with already published results. While comparing varying and 24 hour prediction window datasets for C-Class flares, we observed a higher importance for C-Class flare decay value in varying prediction window dataset. This could be concluded as a result of the lower prediction window at solar maximum letting the model learn only the corresponding samples that affects the observed flare. Apart from this, both the datasets displayed the same top features even though the order was slightly different. For the M Class dataset, the feature importance is similar to that we see for C-Class, except for the fact that MDEC and MHIS showed a higher importance. To properly observe the importance of selecting the right features, we ran the classifier with only the top five features. This confirmed that if we remove the less useful features from the feature set, the model could display a better performance when it comes to flare prediction.\\

We also compared our results with other similar solar flare prediction models, assessing their relative performance through a Taylor diagram based on their optimal TSS and Acc values. Our LightGBM-based model demonstrated better performance over various other models, especially those employing LSTM, RF, MLP, and DeFN, when a limited but highly significant set of features was used. However, the models presented in \cite{nishizuka_2017_solar} yielded better results than ours. Notably, they utilised a more extensive set of features—65 in total—compared to other models in the literature. Despite this, multiple studies, including our own, suggest that focusing on a subset of highly significant features can improve predictive accuracy. This observation opens up an intriguing question regarding the optimal number of features for solar flare prediction: is a larger or smaller set more effective? To address this question comprehensively, further analysis involving model comparisons and in-depth feature selection assessments is required.\\

In the future, with more magnetic data of multiple solar cycles, we aim to improve the performance of the classifier as well as perform more comprehensive study on relative significance of features. During our trial runs, we noticed that instead of using consecutive time frames for training, validating, and testing datasets, if we randomly split the whole data into these datasets, the performance of the models is much better. Therefore, by using solar magnetic data from multiple cycles, we may achieve better accuracy with the solar flare prediction model.


\begin{acks}
 We extend our sincere gratitude to the anonymous referee for the insightful comments and constructive suggestions, which have significantly enhanced the quality of this manuscript. We also thank the editor for the valuable assistance in improving the language of the paper. PAV and PM would like to express their gratitude to Dr. Bhargav Vaidya for his dedicated and insightful discussions. PM would like to acknowledge the ﬁnancial support provided by the Prime Minister’s Research Fellowship.
\end{acks}

%
%
 \bibliographystyle{spr-mp-sola}
 \bibliography{sola_bibliography}  

\begin{thebibliography}{47}
\ifx\bisbn     \undefined \def\bisbn  #1{ISBN #1}\fi
\ifx\binits    \undefined \def\binits#1{#1}\fi
\ifx\bauthor   \undefined \def\bauthor#1{#1}\fi
\ifx\batitle   \undefined \def\batitle#1{#1}\fi
\ifx\bjtitle   \undefined \def\bjtitle#1{\textit{#1}}\fi
\ifx\bvolume   \undefined \def\bvolume#1{\textbf{#1}}\fi
\ifx\byear     \undefined \def\byear#1{#1}\fi
\ifx\bissue    \undefined \def\bissue#1{#1}\fi
\ifx\bfpage    \undefined \def\bfpage#1{#1}\fi
\ifx\blpage    \undefined \def\blpage #1{#1}\fi
\ifx\burl      \undefined \def\burl#1{#1}\fi
\ifx\href      \undefined \def\href#1#2{#2}\fi
\ifx\betal     \undefined \def\betal{et al.}\fi
\ifx\bctitle   \undefined \def\bctitle#1{#1}\fi
\ifx\beditor   \undefined \def\beditor#1{#1}\fi
\ifx\bbtitle   \undefined \def\bbtitle#1{\textit{#1}}\fi
\ifx\bedition  \undefined \def\bedition#1{#1}\fi
\ifx\bseriesno \undefined \def\bseriesno#1{\textbf{#1}}\fi
\ifx\blocation \undefined \def\blocation#1{#1}\fi
\ifx\bsertitle \undefined \def\bsertitle#1{\textit{#1}}\fi
\ifx\bsnm      \undefined \def\bsnm#1{#1}\fi
\ifx\bsuffix   \undefined \def\bsuffix#1{#1}\fi
\ifx\bparticle \undefined \def\bparticle#1{#1}\fi
\ifx\barticle  \undefined \def\barticle#1{}\fi
\ifx\binstitute  \undefined \def\binstitute#1{#1}\fi
\ifx\bpublisher  \undefined \def\bpublisher#1{#1}\fi
\ifx\doiurl    \undefined \def\doiurl#1{\href{#1}{DOI}}\fi
\makeatletter
\def\safeHref#1#2#3{\in@{http}{#2}\ifin@\href{#2}{#3}\else\href{#1#2}{#3}\fi}
\makeatother
\ifx\adsurl    \undefined \def\adsurl#1{\safeHref{https://ui.adsabs.harvard.edu/abs/}{#1}{ADS}}\fi
\ifx\arxivurl  \undefined \def\arxivurl#1{\safeHref{http://arxiv.org/abs/}{#1}{arXiv}}\fi
\ifx\botherref \undefined \def\botherref#1{}\fi
\ifx\url       \undefined \def\url#1{#1}\fi
\ifx\bchapter  \undefined \def\bchapter#1{}\fi
\ifx\bbook     \undefined \def\bbook#1{}\fi
\ifx\bcomment  \undefined \def\bcomment#1{#1}\fi
\ifx\oauthor   \undefined \def\oauthor#1{#1}\fi
\ifx\citeauthoryear \undefined\def \citeauthoryear#1{#1}\fi
\def\endbibitem {}
\ifx\bconflocation  \undefined \def\bconflocation#1{#1} \fi

\bibitem[\protect\citeauthoryear{Ahmadzadeh et~al.}{2019}]{ahmadzadeh2019challenges}
\begin{bchapter}
\bauthor{\bsnm{Ahmadzadeh}, \binits{A.}},
\bauthor{\bsnm{Hostetter}, \binits{M.}},
\bauthor{\bsnm{Aydin}, \binits{B.}},
\bauthor{\bsnm{Georgoulis}, \binits{M.K.}},
\bauthor{\bsnm{Kempton}, \binits{D.J.}},
\bauthor{\bsnm{Mahajan}, \binits{S.S.}},
\bauthor{\bsnm{Angryk}, \binits{R.}}:
\byear{2019},
\bctitle{Challenges with extreme class-imbalance and temporal coherence: A study on solar flare data}.
In: \bbtitle{2019 IEEE international conference on big data (Big Data)},
\bfpage{1423}.
\bcomment{Ieee}.
\end{bchapter}
\endbibitem

\bibitem[\protect\citeauthoryear{Arge and Pizzo}{2000}]{arge_2000_improvement}
\begin{barticle}
\bauthor{\bsnm{Arge}, \binits{O.N.}},
\bauthor{\bsnm{Pizzo}, \binits{V.J.}}:
\byear{2000},
\batitle{Improvement in the prediction of solar wind conditions using near-real time solar magnetic field updates}.
\bjtitle{Journal of Geophysical Research: Space Physics}
\bvolume{105}.
\doiurl{https://doi.org/10.1029/1999ja000262}.
\end{barticle}
\endbibitem

\bibitem[\protect\citeauthoryear{Bloomfield et~al.}{2012}]{Bloomfield_2012}
\begin{barticle}
\bauthor{\bsnm{Bloomfield}, \binits{D.S.}},
\bauthor{\bsnm{Higgins}, \binits{P.A.}},
\bauthor{\bsnm{McAteer}, \binits{R.J.}},
\bauthor{\bsnm{Gallagher}, \binits{P.T.}}:
\byear{2012},
\batitle{Toward reliable benchmarking of solar flare forecasting methods}.
\bjtitle{The Astrophysical Journal Letters}
\bvolume{747},
\bfpage{L41}.
\doiurl{https://doi.org/10.1088/2041-8205/747/2/l41}.
\end{barticle}
\endbibitem

\bibitem[\protect\citeauthoryear{Bobra and Couvidat}{2015}]{bobra_2015_solar}
\begin{barticle}
\bauthor{\bsnm{Bobra}, \binits{M.G.}},
\bauthor{\bsnm{Couvidat}, \binits{S.}}:
\byear{2015},
\batitle{Solar flare prediction using SDO/HMI vector magnetic field data with a machine-learning algorithm}.
\bjtitle{The Astrophysical Journal}
\bvolume{798},
\bfpage{135}.
\doiurl{https://doi.org/10.1088/0004-637x/798/2/135}.
\end{barticle}
\endbibitem

\bibitem[\protect\citeauthoryear{Bobra et~al.}{2014}]{Bobra_2014}
\begin{barticle}
\bauthor{\bsnm{Bobra}, \binits{M.G.}},
\bauthor{\bsnm{Sun}, \binits{X.}},
\bauthor{\bsnm{Hoeksema}, \binits{J.T.}},
\bauthor{\bsnm{Turmon}, \binits{M.}},
\bauthor{\bsnm{Liu}, \binits{Y.}},
\bauthor{\bsnm{Hayashi}, \binits{K.}},
\bauthor{\bsnm{Barnes}, \binits{G.}},
\bauthor{\bsnm{Leka}, \binits{K.D.}}:
\byear{2014},
\batitle{The Helioseismic and Magnetic Imager (HMI) vector magnetic field pipeline: SHARPs – Space-Weather HMI Active Region Patches}.
\bjtitle{Solar Physics}
\bvolume{289},
\bfpage{3549–3578}.
\doiurl{https://doi.org/10.1007/s11207-014-0529-3}.
\end{barticle}
\endbibitem

\bibitem[\protect\citeauthoryear{Bussy-Virat and Ridley}{2014}]{bussyvirat_2014_predictions}
\begin{barticle}
\bauthor{\bsnm{Bussy-Virat}, \binits{C.D.}},
\bauthor{\bsnm{Ridley}, \binits{A.J.}}:
\byear{2014},
\batitle{Predictions of the solar wind speed by the probability distribution function model}.
\bjtitle{Space Weather}
\bvolume{12}.
\doiurl{https://doi.org/10.1002/2014SW001051}.
\end{barticle}
\endbibitem

\bibitem[\protect\citeauthoryear{Chawla et~al.}{2002}]{Chawla_2002}
\begin{barticle}
\bauthor{\bsnm{Chawla}, \binits{N.V.}},
\bauthor{\bsnm{Bowyer}, \binits{K.W.}},
\bauthor{\bsnm{Hall}, \binits{L.O.}},
\bauthor{\bsnm{Kegelmeyer}, \binits{W.P.}}:
\byear{2002},
\batitle{SMOTE: Synthetic Minority Over-sampling Technique}.
\bjtitle{Journal of Artificial Intelligence Research}
\bvolume{16},
\bfpage{321–357}.
\doiurl{https://doi.org/10.1613/jair.953}.
\end{barticle}
\endbibitem

\bibitem[\protect\citeauthoryear{Chen et~al.}{2019}]{chen_2019_identifying}
\begin{barticle}
\bauthor{\bsnm{Chen}, \binits{Y.}},
\bauthor{\bsnm{Manchester}, \binits{W.B.}},
\bauthor{\bsnm{Hero}, \binits{A.O.}},
\bauthor{\bsnm{Toth}, \binits{G.}},
\bauthor{\bsnm{DuFumier}, \binits{B.}},
\bauthor{\bsnm{Zhou}, \binits{T.}},
\bauthor{\bsnm{Wang}, \binits{X.}},
\bauthor{\bsnm{Zhu}, \binits{H.}},
\bauthor{\bsnm{Sun}, \binits{Z.}},
\bauthor{\bsnm{Gombosi}, \binits{T.I.}}:
\byear{2019},
\batitle{Identifying solar flare precursors using time series of SDO/HMI Images and SHARP parameters}.
\bjtitle{Space Weather}
\bvolume{17},
\bfpage{1404}.
\doiurl{https://doi.org/10.1029/2019sw002214}.
\end{barticle}
\endbibitem

\bibitem[\protect\citeauthoryear{Community et~al.}{2015}]{Community_2015}
\begin{barticle}
\bauthor{\bsnm{Community}, \binits{T.S.}},
\bauthor{\bsnm{Mumford}, \binits{S.J.}},
\bauthor{\bsnm{Christe}, \binits{S.}},
\bauthor{\bsnm{P{\'{e}}rez-Su{\'{a}}rez}, \binits{D.}},
\bauthor{\bsnm{Ireland}, \binits{J.}},
\bauthor{\bsnm{Shih}, \binits{A.Y.}},
\bauthor{\bsnm{Inglis}, \binits{A.R.}},
\bauthor{\bsnm{Liedtke}, \binits{S.}},
\bauthor{\bsnm{Hewett}, \binits{R.J.}},
\bauthor{\bsnm{Mayer}, \binits{F.}},
\bauthor{\bsnm{Hughitt}, \binits{K.}},
\bauthor{\bsnm{Freij}, \binits{N.}},
\bauthor{\bsnm{Meszaros}, \binits{T.}},
\bauthor{\bsnm{Bennett}, \binits{S.M.}},
\bauthor{\bsnm{Malocha}, \binits{M.}},
\bauthor{\bsnm{Evans}, \binits{J.}},
\bauthor{\bsnm{Agrawal}, \binits{A.}},
\bauthor{\bsnm{Leonard}, \binits{A.J.}},
\bauthor{\bsnm{Robitaille}, \binits{T.P.}},
\bauthor{\bsnm{Mampaey}, \binits{B.}},
\bauthor{\bsnm{Campos-Rozo}, \binits{J.I.}},
\bauthor{\bsnm{Kirk}, \binits{M.S.}}:
\byear{2015},
\batitle{{SunPy}{\textemdash}Python for solar physics}.
\bjtitle{Computational Science {\&} Discovery}
\bvolume{8},
\bfpage{014009}.
\doiurl{https://doi.org/10.1088/1749-4699/8/1/014009}.
\end{barticle}
\endbibitem

\bibitem[\protect\citeauthoryear{Echer et~al.}{2005}]{ECHER2005855}
\begin{barticle}
\bauthor{\bsnm{Echer}, \binits{E.}},
\bauthor{\bsnm{Gonzalez}, \binits{W.D.}},
\bauthor{\bsnm{Guarnieri}, \binits{F.L.}},
\bauthor{\bsnm{Lago}, \binits{A.D.}},
\bauthor{\bsnm{Vieira}, \binits{L.E.A.}}:
\byear{2005},
\batitle{Introduction to space weather}.
\bjtitle{Advances in Space Research}
\bvolume{35},
\bfpage{855}.
\bcomment{Fundamentals of Space Environment Science}.
\doiurl{https://doi.org/10.1016/j.asr.2005.02.098}.
\end{barticle}
\endbibitem

\bibitem[\protect\citeauthoryear{Falconer et~al.}{2012}]{Falconer_2012}
\begin{barticle}
\bauthor{\bsnm{Falconer}, \binits{D.A.}},
\bauthor{\bsnm{Moore}, \binits{R.L.}},
\bauthor{\bsnm{Barghouty}, \binits{A.F.}},
\bauthor{\bsnm{Khazanov}, \binits{I.}}:
\byear{2012},
\batitle{Prior flaring as a complement to free magnetic energy for forecasting solar eruptions}.
\bjtitle{The Astrophysical Journal}
\bvolume{757},
\bfpage{32}.
\doiurl{https://doi.org/10.1088/0004-637x/757/1/32}.
\end{barticle}
\endbibitem

\bibitem[\protect\citeauthoryear{Fisher et~al.}{2011}]{fisher_2011_global}
\begin{barticle}
\bauthor{\bsnm{Fisher}, \binits{G.H.}},
\bauthor{\bsnm{Bercik}, \binits{D.J.}},
\bauthor{\bsnm{Welsch}, \binits{B.T.}},
\bauthor{\bsnm{Hudson}, \binits{H.S.}}:
\byear{2011},
\batitle{Global forces in eruptive solar flares: The Lorentz force acting on the solar atmosphere and the solar interior}.
\bjtitle{Solar Physics}
\bvolume{277},
\bfpage{59}.
\doiurl{https://doi.org/10.1007/s11207-011-9907-2}.
\end{barticle}
\endbibitem

\bibitem[\protect\citeauthoryear{Florios et~al.}{2018}]{florios_2018_forecasting}
\begin{barticle}
\bauthor{\bsnm{Florios}, \binits{K.}},
\bauthor{\bsnm{Kontogiannis}, \binits{I.}},
\bauthor{\bsnm{Park}, \binits{S.-H.}},
\bauthor{\bsnm{Guerra}, \binits{J.A.}},
\bauthor{\bsnm{Benvenuto}, \binits{F.}},
\bauthor{\bsnm{Bloomfield}, \binits{D.S.}},
\bauthor{\bsnm{Georgoulis}, \binits{M.K.}}:
\byear{2018},
\batitle{Forecasting solar flares using magnetogram-based predictors and machine learning}.
\bjtitle{Solar Physics}
\bvolume{293}.
\doiurl{https://doi.org/10.1007/s11207-018-1250-4}.
\end{barticle}
\endbibitem

\bibitem[\protect\citeauthoryear{Han, Wang, and Mao}{2005}]{10.1007/11538059_91}
\begin{bchapter}
\bauthor{\bsnm{Han}, \binits{H.}},
\bauthor{\bsnm{Wang}, \binits{W.-Y.}},
\bauthor{\bsnm{Mao}, \binits{B.-H.}}:
\byear{2005},
\bctitle{Borderline-SMOTE: a new over-sampling method in imbalanced data sets learning}.
In: \bbtitle{International conference on intelligent computing},
\bfpage{878}.
\bcomment{Springer}.
\end{bchapter}
\endbibitem

\bibitem[\protect\citeauthoryear{He and Garcia}{2009}]{5128907}
\begin{barticle}
\bauthor{\bsnm{He}, \binits{H.}},
\bauthor{\bsnm{Garcia}, \binits{E.A.}}:
\byear{2009},
\batitle{Learning from imbalanced data}.
\bjtitle{IEEE Transactions on Knowledge and Data Engineering}
\bvolume{21},
\bfpage{1263}.
\doiurl{https://doi.org/10.1109/TKDE.2008.239}.
\end{barticle}
\endbibitem

\bibitem[\protect\citeauthoryear{He et~al.}{2008}]{4633969}
\begin{bchapter}
\bauthor{\bsnm{He}, \binits{H.}},
\bauthor{\bsnm{Bai}, \binits{Y.}},
\bauthor{\bsnm{Garcia}, \binits{E.A.}},
\bauthor{\bsnm{Li}, \binits{S.}}:
\byear{2008},
\bctitle{ADASYN: Adaptive synthetic sampling approach for imbalanced learning}.
In: \bbtitle{2008 IEEE International Joint Conference on Neural Networks (IEEE World Congress on Computational Intelligence)},
\bfpage{1322}.
\doiurl{https://doi.org/10.1109/IJCNN.2008.4633969}.
\end{bchapter}
\endbibitem

\bibitem[\protect\citeauthoryear{Huang and Wang}{2013}]{huang_2013_solar}
\begin{barticle}
\bauthor{\bsnm{Huang}, \binits{X.}},
\bauthor{\bsnm{Wang}, \binits{H.-N.}}:
\byear{2013},
\batitle{Solar flare prediction using highly stressed longitudinal magnetic field parameters}.
\bjtitle{Research in Astronomy and Astrophysics}
\bvolume{13},
\bfpage{351}.
\doiurl{https://doi.org/10.1088/1674-4527/13/3/010}.
\end{barticle}
\endbibitem

\bibitem[\protect\citeauthoryear{Jiao et~al.}{2020}]{jiao_2020_solar}
\begin{barticle}
\bauthor{\bsnm{Jiao}, \binits{Z.}},
\bauthor{\bsnm{Sun}, \binits{H.}},
\bauthor{\bsnm{Wang}, \binits{X.}},
\bauthor{\bsnm{Manchester}, \binits{W.}},
\bauthor{\bsnm{Gombosi}, \binits{T.}},
\bauthor{\bsnm{Hero}, \binits{A.}},
\bauthor{\bsnm{Chen}, \binits{Y.}}:
\byear{2020},
\batitle{Solar flare intensity prediction with machine learning models}.
\bjtitle{Space Weather}
\bvolume{18}.
\doiurl{https://doi.org/10.1029/2020sw002440}.
\end{barticle}
\endbibitem

\bibitem[\protect\citeauthoryear{Jonas et~al.}{2018}]{Jonas_2018}
\begin{barticle}
\bauthor{\bsnm{Jonas}, \binits{E.}},
\bauthor{\bsnm{Bobra}, \binits{M.}},
\bauthor{\bsnm{Shankar}, \binits{V.}},
\bauthor{\bsnm{Todd~Hoeksema}, \binits{J.}},
\bauthor{\bsnm{Recht}, \binits{B.}}:
\byear{2018},
\batitle{Flare prediction using photospheric and coronal image data}.
\bjtitle{Solar Physics}
\bvolume{293}.
\doiurl{https://doi.org/10.1007/s11207-018-1258-9}.
\end{barticle}
\endbibitem

\bibitem[\protect\citeauthoryear{Ke et~al.}{2017a}]{ke_2017_lightgbm}
\begin{barticle}
\bauthor{\bsnm{Ke}, \binits{G.}},
\bauthor{\bsnm{Meng}, \binits{Q.}},
\bauthor{\bsnm{Finley}, \binits{T.}},
\bauthor{\bsnm{Wang}, \binits{T.}},
\bauthor{\bsnm{Chen}, \binits{W.}},
\bauthor{\bsnm{Ma}, \binits{W.}},
\bauthor{\bsnm{Ye}, \binits{Q.}},
\bauthor{\bsnm{Liu}, \binits{T.-Y.}}:
\byear{2017}a,
\batitle{LightGBM: a highly efficient gradient boosting decision tree}.
\bjtitle{Neural Information Processing Systems}
\bvolume{30},
\bfpage{3149}.
\end{barticle}
\endbibitem

\bibitem[\protect\citeauthoryear{Ke et~al.}{2017b}]{ke2017lightgbm}
\begin{botherref}
\oauthor{\bsnm{Ke}, \binits{G.}},
\oauthor{\bsnm{Meng}, \binits{Q.}},
\oauthor{\bsnm{Finley}, \binits{T.}},
\oauthor{\bsnm{Wang}, \binits{T.}},
\oauthor{\bsnm{Chen}, \binits{W.}},
\oauthor{\bsnm{Ma}, \binits{W.}},
\oauthor{\bsnm{Ye}, \binits{Q.}},
\oauthor{\bsnm{Liu}, \binits{T.-Y.}}:
2017b,
Lightgbm: A highly efficient gradient boosting decision tree.
\textit{Advances in neural information processing systems}
\textbf{30}.
\end{botherref}
\endbibitem

\bibitem[\protect\citeauthoryear{LaBonte, Georgoulis, and Rust}{2007}]{labonte_2007_survey}
\begin{barticle}
\bauthor{\bsnm{LaBonte}, \binits{B.J.}},
\bauthor{\bsnm{Georgoulis}, \binits{M.K.}},
\bauthor{\bsnm{Rust}, \binits{D.M.}}:
\byear{2007},
\batitle{Survey of magnetic helicity injection in regions producing X‐Class flares}.
\bjtitle{The Astrophysical Journal}
\bvolume{671},
\bfpage{955}.
\doiurl{https://doi.org/10.1086/522682}.
\end{barticle}
\endbibitem

\bibitem[\protect\citeauthoryear{Leka and Barnes}{2003}]{leka_2003_photospheric}
\begin{barticle}
\bauthor{\bsnm{Leka}, \binits{K.D.}},
\bauthor{\bsnm{Barnes}, \binits{G.}}:
\byear{2003},
\batitle{Photospheric magnetic field properties of flaring versus flare-quiet active regions. II. Discriminant analysis}.
\bjtitle{The Astrophysical Journal}
\bvolume{595},
\bfpage{1296}.
\doiurl{https://doi.org/10.1086/377512}.
\end{barticle}
\endbibitem

\bibitem[\protect\citeauthoryear{Li and Zhu}{2013}]{li_2013_solar}
\begin{barticle}
\bauthor{\bsnm{Li}, \binits{R.}},
\bauthor{\bsnm{Zhu}, \binits{J.}}:
\byear{2013},
\batitle{Solar flare forecasting based on sequential sunspot data}.
\bjtitle{Research in Astronomy and Astrophysics}
\bvolume{13},
\bfpage{1118}.
\doiurl{https://doi.org/10.1088/1674-4527/13/9/010}.
\end{barticle}
\endbibitem

\bibitem[\protect\citeauthoryear{Lin et~al.}{2017}]{Lin_2017_ICCV}
\begin{bchapter}
\bauthor{\bsnm{Lin}, \binits{T.-Y.}},
\bauthor{\bsnm{Goyal}, \binits{P.}},
\bauthor{\bsnm{Girshick}, \binits{R.}},
\bauthor{\bsnm{He}, \binits{K.}},
\bauthor{\bsnm{Dollar}, \binits{P.}}:
\byear{2017},
\bctitle{Focal loss for dense object detection}.
In: \bbtitle{Proceedings of the IEEE International Conference on Computer Vision (ICCV)}.
\end{bchapter}
\endbibitem

\bibitem[\protect\citeauthoryear{Liu et~al.}{2019}]{liu_2019_predicting}
\begin{barticle}
\bauthor{\bsnm{Liu}, \binits{H.}},
\bauthor{\bsnm{Liu}, \binits{C.}},
\bauthor{\bsnm{Wang}, \binits{J.T.L.}},
\bauthor{\bsnm{Wang}, \binits{H.}}:
\byear{2019},
\batitle{Predicting solar flares using a long short-term memory network}.
\bjtitle{The Astrophysical Journal}
\bvolume{877},
\bfpage{121}.
\doiurl{https://doi.org/10.3847/1538-4357/ab1b3c}.
\end{barticle}
\endbibitem

\bibitem[\protect\citeauthoryear{Liu et~al.}{2017}]{liu_2017_shortterm}
\begin{barticle}
\bauthor{\bsnm{Liu}, \binits{J.-F.}},
\bauthor{\bsnm{Li}, \binits{F.}},
\bauthor{\bsnm{Wan}, \binits{J.}},
\bauthor{\bsnm{Yu}, \binits{D.-R.}}:
\byear{2017},
\batitle{Short-term solar flare prediction using multi-model integration method}.
\bjtitle{Research in Astronomy and Astrophysics}
\bvolume{17},
\bfpage{034}.
\doiurl{https://doi.org/10.1088/1674-4527/17/4/34}.
\end{barticle}
\endbibitem

\bibitem[\protect\citeauthoryear{Liu et~al.}{2023}]{Liu_2023}
\begin{barticle}
\bauthor{\bsnm{Liu}, \binits{Y.}},
\bauthor{\bsnm{Welsch}, \binits{B.T.}},
\bauthor{\bsnm{Valori}, \binits{G.}},
\bauthor{\bsnm{Georgoulis}, \binits{M.K.}},
\bauthor{\bsnm{Guo}, \binits{Y.}},
\bauthor{\bsnm{Pariat}, \binits{E.}},
\bauthor{\bsnm{Park}, \binits{S.-H.}},
\bauthor{\bsnm{Thalmann}, \binits{J.K.}}:
\byear{2023},
\batitle{Changes of magnetic energy and helicity in solar active regions from major flares}.
\bjtitle{The Astrophysical Journal}
\bvolume{942},
\bfpage{27}.
\doiurl{https://doi.org/10.3847/1538-4357/aca3a6}.
\end{barticle}
\endbibitem

\bibitem[\protect\citeauthoryear{Mayank, Vaidya, and Chakrabarty}{2022}]{mayank_2022_swastisw}
\begin{barticle}
\bauthor{\bsnm{Mayank}, \binits{P.}},
\bauthor{\bsnm{Vaidya}, \binits{B.}},
\bauthor{\bsnm{Chakrabarty}, \binits{D.}}:
\byear{2022},
\batitle{SWASTi-SW: Space weather adaptive simulation framework for solar wind and its relevance to the Aditya-L1 mission}.
\bjtitle{The Astrophysical Journal Supplement Series}
\bvolume{262},
\bfpage{23}.
\doiurl{https://doi.org/10.3847/1538-4365/ac8551}.
\end{barticle}
\endbibitem

\bibitem[\protect\citeauthoryear{Moore, Falconer, and Sterling}{2012}]{moore_2012_the}
\begin{barticle}
\bauthor{\bsnm{Moore}, \binits{R.L.}},
\bauthor{\bsnm{Falconer}, \binits{D.A.}},
\bauthor{\bsnm{Sterling}, \binits{A.C.}}:
\byear{2012},
\batitle{The limit of magnetic-shear energy in solar active regions}.
\bjtitle{The Astrophysical Journal}
\bvolume{750},
\bfpage{24}.
\doiurl{https://doi.org/10.1088/0004-637x/750/1/24}.
\end{barticle}
\endbibitem

\bibitem[\protect\citeauthoryear{Nguyen, Cooper, and Kamei}{2011}]{nguyen2011borderline}
\begin{barticle}
\bauthor{\bsnm{Nguyen}, \binits{H.M.}},
\bauthor{\bsnm{Cooper}, \binits{E.W.}},
\bauthor{\bsnm{Kamei}, \binits{K.}}:
\byear{2011},
\batitle{Borderline over-sampling for imbalanced data classification}.
\bjtitle{International Journal of Knowledge Engineering and Soft Data Paradigms}
\bvolume{3},
\bfpage{4}.
\end{barticle}
\endbibitem

\bibitem[\protect\citeauthoryear{Nishizuka et~al.}{2017}]{nishizuka_2017_solar}
\begin{barticle}
\bauthor{\bsnm{Nishizuka}, \binits{N.}},
\bauthor{\bsnm{Sugiura}, \binits{K.}},
\bauthor{\bsnm{Kubo}, \binits{Y.}},
\bauthor{\bsnm{Den}, \binits{M.}},
\bauthor{\bsnm{Watari}, \binits{S.}},
\bauthor{\bsnm{Ishii}, \binits{M.}}:
\byear{2017},
\batitle{Solar Flare Prediction Model with Three Machine-learning Algorithms using Ultraviolet Brightening and Vector Magnetograms}.
\bjtitle{The Astrophysical Journal}
\bvolume{835},
\bfpage{156}.
\doiurl{https://doi.org/10.3847/1538-4357/835/2/156}.
\end{barticle}
\endbibitem

\bibitem[\protect\citeauthoryear{Nishizuka et~al.}{2018}]{nishizuka_2018_deep}
\begin{barticle}
\bauthor{\bsnm{Nishizuka}, \binits{N.}},
\bauthor{\bsnm{Sugiura}, \binits{K.}},
\bauthor{\bsnm{Kubo}, \binits{Y.}},
\bauthor{\bsnm{Den}, \binits{M.}},
\bauthor{\bsnm{Ishii}, \binits{M.}}:
\byear{2018},
\batitle{Deep Flare Net (DeFN) model for solar flare prediction}.
\bjtitle{The Astrophysical Journal}
\bvolume{858},
\bfpage{113}.
\doiurl{https://doi.org/10.3847/1538-4357/aab9a7}.
\end{barticle}
\endbibitem

\bibitem[\protect\citeauthoryear{Odstrcil}{2003}]{dodstrcil_2003_modeling}
\begin{barticle}
\bauthor{\bsnm{Odstrcil}, \binits{D.}}:
\byear{2003},
\batitle{Modeling 3-D solar wind structure}.
\bjtitle{Advances in Space Research}
\bvolume{32},
\bfpage{497}.
\doiurl{https://doi.org/10.1016/S0273-1177(03)00332-6}.
\end{barticle}
\endbibitem

\bibitem[\protect\citeauthoryear{Oughton et~al.}{2019}]{oughton_2019_a}
\begin{barticle}
\bauthor{\bsnm{Oughton}, \binits{E.J.}},
\bauthor{\bsnm{Hapgood}, \binits{M.}},
\bauthor{\bsnm{Richardson}, \binits{G.}},
\bauthor{\bsnm{Beggan}, \binits{C.}},
\bauthor{, \binits{A.}},
\bauthor{\bsnm{Gibbs}, \binits{M.T.}},
\bauthor{\bsnm{Burnett}, \binits{C.}},
\bauthor{\bsnm{Gaunt}, \binits{C.T.}},
\bauthor{\bsnm{Trichas}, \binits{M.}},
\bauthor{\bsnm{Dada}, \binits{R.}},
\bauthor{\bsnm{Horne}, \binits{R.B.}}:
\byear{2019},
\batitle{A risk assessment framework for the socioeconomic impacts of electricity transmission infrastructure failure due to space weather: An application to the United Kingdom}.
\bjtitle{Risk Analysis}
\bvolume{39},
\bfpage{1022}.
\doiurl{https://doi.org/10.1111/risa.13229}.
\end{barticle}
\endbibitem

\bibitem[\protect\citeauthoryear{Owens, Riley, and Horbury}{2017}]{owens_2017_probabilistic}
\begin{barticle}
\bauthor{\bsnm{Owens}, \binits{M.J.}},
\bauthor{\bsnm{Riley}, \binits{P.}},
\bauthor{\bsnm{Horbury}, \binits{T.S.}}:
\byear{2017},
\batitle{Probabilistic solar wind and geomagnetic forecasting using an analogue ensemble or “Similar day” approach}.
\bjtitle{Solar Physics}
\bvolume{292}.
\doiurl{https://doi.org/10.1007/s11207-017-1090-7}.
\end{barticle}
\endbibitem

\bibitem[\protect\citeauthoryear{Park, Chae, and Wang}{2010}]{Park_2010}
\begin{barticle}
\bauthor{\bsnm{Park}, \binits{S.-h.}},
\bauthor{\bsnm{Chae}, \binits{J.}},
\bauthor{\bsnm{Wang}, \binits{H.}}:
\byear{2010},
\batitle{Productivity of solar flares and magnetic helicity injection in active regions}.
\bjtitle{The Astrophysical Journal}
\bvolume{718},
\bfpage{43}.
\doiurl{https://doi.org/10.1088/0004-637X/718/1/43}.
\end{barticle}
\endbibitem

\bibitem[\protect\citeauthoryear{Pomoell and Poedts}{2018}]{jenspomoell_2018_euhforia}
\begin{barticle}
\bauthor{\bsnm{Pomoell}, \binits{J.}},
\bauthor{\bsnm{Poedts}, \binits{S.}}:
\byear{2018},
\batitle{EUHFORIA: European heliospheric forecasting information asset}.
\bjtitle{Journal of Space Weather and Space Climate}
\bvolume{8},
\bfpage{A35}.
\doiurl{https://doi.org/10.1051/swsc/2018020}.
\end{barticle}
\endbibitem

\bibitem[\protect\citeauthoryear{Reiss et~al.}{2016}]{reiss_2016_verification}
\begin{barticle}
\bauthor{\bsnm{Reiss}, \binits{M.A.}},
\bauthor{\bsnm{Temmer}, \binits{M.}},
\bauthor{\bsnm{Veronig}, \binits{A.M.}},
\bauthor{\bsnm{Nikolic}, \binits{L.}},
\bauthor{\bsnm{Vennerstrom}, \binits{S.}},
\bauthor{\bsnm{Schöngassner}, \binits{F.}},
\bauthor{\bsnm{Hofmeister}, \binits{S.J.}}:
\byear{2016},
\batitle{Verification of high-speed solar wind stream forecasts using operational solar wind models}.
\bjtitle{Space Weather}
\bvolume{14}.
\doiurl{https://doi.org/10.1002/2016SW001390}.
\end{barticle}
\endbibitem

\bibitem[\protect\citeauthoryear{Ribeiro and Gradvohl}{2021}]{ribeiro_2021_machine}
\begin{barticle}
\bauthor{\bsnm{Ribeiro}, \binits{F.}},
\bauthor{\bsnm{Gradvohl}, \binits{A.L.S.}}:
\byear{2021},
\batitle{Machine learning techniques applied to solar flares forecasting}.
\bjtitle{Astronomy and Computing}
\bvolume{35},
\bfpage{100468}.
\doiurl{https://doi.org/10.1016/j.ascom.2021.100468}.
\end{barticle}
\endbibitem

\bibitem[\protect\citeauthoryear{Riley, Linker, and Mikić}{2001}]{riley_2001_an}
\begin{barticle}
\bauthor{\bsnm{Riley}, \binits{P.}},
\bauthor{\bsnm{Linker}, \binits{J.A.}},
\bauthor{\bsnm{Mikić}, \binits{Z.}}:
\byear{2001},
\batitle{An empirically-driven global MHD model of the solar corona and inner heliosphere}.
\bjtitle{Journal of Geophysical Research: Space Physics}
\bvolume{106},
\bfpage{15889}.
\doiurl{https://doi.org/10.1029/2000JA000121}.
\end{barticle}
\endbibitem

\bibitem[\protect\citeauthoryear{Riley et~al.}{2017}]{riley_2017_forecasting}
\begin{barticle}
\bauthor{\bsnm{Riley}, \binits{P.}},
\bauthor{\bsnm{Ben-Nun}, \binits{M.}},
\bauthor{\bsnm{Linker}, \binits{J.A.}},
\bauthor{\bsnm{Owens}, \binits{M.J.}},
\bauthor{\bsnm{Horbury}, \binits{T.S.}}:
\byear{2017},
\batitle{Forecasting the properties of the solar wind using simple pattern recognition}.
\bjtitle{Space Weather}
\bvolume{15}.
\doiurl{https://doi.org/10.1002/2016SW001589}.
\end{barticle}
\endbibitem

\bibitem[\protect\citeauthoryear{Schrijver}{2007}]{schrijver_2007_a}
\begin{barticle}
\bauthor{\bsnm{Schrijver}, \binits{C.J.}}:
\byear{2007},
\batitle{A characteristic magnetic field pattern associated with all major solar flares and its use in flare forecasting}.
\bjtitle{The Astrophysical Journal}
\bvolume{655},
\bfpage{L117–L120}.
\doiurl{https://doi.org/10.1086/511857}.
\end{barticle}
\endbibitem

\bibitem[\protect\citeauthoryear{Sun}{2019}]{sun2019cgem}
\begin{botherref}
\oauthor{\bsnm{Sun}, \binits{X.}}:
2019,
\textit{The CGEM Lorentz Force Data from HMI Vector Magnetograms}.
\end{botherref}
\endbibitem

\bibitem[\protect\citeauthoryear{Taylor}{2001}]{taylor2001summarizing}
\begin{barticle}
\bauthor{\bsnm{Taylor}, \binits{K.E.}}:
\byear{2001},
\batitle{Summarizing multiple aspects of model performance in a single diagram}.
\bjtitle{Journal of Geophysical Research: Atmospheres}
\bvolume{106},
\bfpage{7183}.
\end{barticle}
\endbibitem

\bibitem[\protect\citeauthoryear{T{\'{o}}th, van~der Holst, and Huang}{2011}]{gbortth_2011_obtaining}
\begin{barticle}
\bauthor{\bsnm{T{\'{o}}th}, \binits{G.}},
\bauthor{\bparticle{van~der} \bsnm{Holst}, \binits{B.}},
\bauthor{\bsnm{Huang}, \binits{Z.}}:
\byear{2011},
\batitle{Obtaining potential field solutions with spherical harmonics and finite differences}.
\bjtitle{The Astrophysical Journal}
\bvolume{732},
\bfpage{102}.
\doiurl{https://doi.org/10.1088/0004-637x/732/2/102}.
\end{barticle}
\endbibitem

\bibitem[\protect\citeauthoryear{Wang et~al.}{2020}]{wang_2020_predicting}
\begin{barticle}
\bauthor{\bsnm{Wang}, \binits{X.}},
\bauthor{\bsnm{Chen}, \binits{Y.}},
\bauthor{\bsnm{Toth}, \binits{G.}},
\bauthor{\bsnm{Manchester}, \binits{W.B.}},
\bauthor{\bsnm{Gombosi}, \binits{T.I.}},
\bauthor{\bsnm{Hero}, \binits{A.O.}},
\bauthor{\bsnm{Jiao}, \binits{Z.}},
\bauthor{\bsnm{Sun}, \binits{H.}},
\bauthor{\bsnm{Jin}, \binits{M.}},
\bauthor{\bsnm{Liu}, \binits{Y.}}:
\byear{2020},
\batitle{Predicting solar flares with machine learning: Investigating solar cycle dependence}.
\bjtitle{The Astrophysical Journal}
\bvolume{895},
\bfpage{3}.
\doiurl{https://doi.org/10.3847/1538-4357/ab89ac}.
\end{barticle}
\endbibitem

\end{thebibliography}
%
%
%
%

\end{article} 
\end{document}